\documentstyle[seceq,preprint,epsbox,epsf]{jpsj}

\def\runtitle{Effect of Magnetic field on the Pseudogap Phenomena 
in High-$T_{c}$ Cuprates} 
\def\runauthor{Youichi {\sc Yanase}, Kosaku {\sc Yamada} }

\title{ Effect of Magnetic field on the Pseudogap Phenomena \\
in High-$T_{{\rm c}}$ Cuprates }

\author{Youichi {\sc Yanase}\footnote{E-mail: yanase@ton.scphys.kyoto-u.ac.jp}
 and Kosaku {\sc Yamada}}

\inst{Department of Physics, Kyoto University, Kyoto 606-8502}

\recdate{November 1, 1999}

\abst
{
 We theoretically investigate the effect of magnetic field on the pseudogap 
phenomena in High-$T_{{\rm c}}$ cuprates. The obtained results well explain 
the experimental results including their doping dependences. 
 In our previous paper (J. Phys. Soc. Jpn. {\bf68} (1999) 2999.), 
we have shown that the pseudogap phenomena observed in 
High-$T_{{\rm c}}$ cuprates are naturally understood as a precursor of 
the strong coupling superconductivity. 
 On the other hand, there is an interpretation for the recent high field NMR 
measurements to be an evidence denying the pairing scenarios for the pseudogap.
 In this paper, we  investigate the magnetic field dependence of NMR $1/T_{1}T$
on the basis of our formalism 
and show the interpretation to be inappropriate. 
 We consider the Landau quantization for the superconducting fluctuations as a 
main effect of the magnetic field.  
 The results indicate that the value of the characteristic magnetic field 
($B_{{\rm ch}}$) is remarkably large in case of the strong coupling 
superconductivity, especially near the pseudogap onset temperature ($T^{*}$). 
 Therefore, the magnetic field dependences can not 
be observed and $T^{*}$ does not vary when the strong pseudogap anomaly is 
observed. 
 On the other hand, $B_{{\rm ch}}$ is small in the comparatively weak coupling 
case and $T^{*}$ varies when the weak pseudogap phenomena are observed. 
 These results properly explain the high magnetic field NMR experiments 
continuously from under-doped to over-doped cuprates. 
 Moreover, we discuss the transport phenomena in the pseudogap phase. 
The behaviors of the in-plane resistivity, the Hall coefficient and the 
{\it c}-axis resistivity in the pseudogap phase are naturally understood 
by considering the $d_{x^{2}-y^{2}}$-wave pseudogap. 
}

\kword
{high-$T_{{\rm c}}$ cuprates, pseudogap phenomena, 
strong coupling superconductivity, \\
superconducting fluctuation, $1/T_{1}T$, magnetic field dependence, 
transport coefficient} 

\begin{document}
\sloppy
\maketitle

\section{Introduction}

 Since the discovery of high-temperature (High-$T_{{\rm c}}$) 
superconductivity by Bednortz and M$\rm{\ddot{u}}$ller,~\cite{rf:bednortz} 
the anomalous normal state properties have been studied for many years 
from the various points of view. 

 In particular, the pseudogap phenomena in under-doped cuprates have been 
recognized as one of the most important issues. 
 There are enormous studies for the issue from both experimental and 
theoretical points of view. 
 However, the complete understanding still remains to be obtained. 

 The pseudogap phenomena mean the suppression of the spectral weight near the 
Fermi energy without any long range order. 
 They are universal phenomena observed in various compounds of under-doped 
cuprates. 

 Various experiments such as nuclear magnetic resonance (NMR),~\cite{rf:NMR}  
optical conductivity,~\cite{rf:homes} transport,~\cite{rf:transport} 
angle-resolved photo-emission spectroscopy 
(ARPES),~\cite{rf:ARPES} tunneling spectroscopy,~\cite{rf:renner} 
electronic specific heat,~\cite{rf:momono} and so on 
have indicated the existence of the pseudogap in the normal state 
High-$T_{{\rm c}}$ cuprates from optimally-doped to under-doped region. 
 In particular, NMR measurements of $1/T_{1}T$ have shown the existence of 
the pseudogap in the spin excitation channel from early years.~\cite{rf:NMR} 

 In the previous paper, we have explained the pseudogap phenomena 
as a precursor of the strong coupling superconductivity.~\cite{rf:yanasepg} 
 Since the effective Fermi energy $\varepsilon_{{\rm F}}$ is renormalized by 
the electron-electron correlation, the ratio 
$T_{{\rm c}}/\varepsilon_{{\rm F}}$ increases in the strongly correlated 
electron systems. The ratio indicates the strength of the superconducting 
coupling. 
 Therefore, the strong coupling superconductivity has a general importance 
for the superconductivity in the strongly correlated electron systems. 
 Moreover, it is natural to consider the strong coupling superconductivity 
in High-$T_{{\rm c}}$ cuprates because of the high critical temperature 
$T_{{\rm c}}$ itself. 
 The strong coupling superconductivity necessarily leads to the strong thermal 
superconducting fluctuations. Such strong fluctuations in the quasi-two 
dimensional systems have serious effects on the electronic state and give rise 
to the pseudogap phenomena.~\cite{rf:yanasepg} 
 
 Actually, various experiments have indicated the close relationship 
between the pseudogap phenomena and the superconductivity. 
 In particular, ARPES have 
directly shown the pseudogap in the one-particle spectral 
weight~\cite{rf:ARPES} and suggested its close relevance and continuity to the 
superconducting gap.~\cite{rf:norman} 

 Other scenarios have been theoretically proposed for the pseudogap phenomena. 
 In the resonating valence bond (RVB) theory, 
there are two distinct excitations, spinon and holon. 
The pseudogap is described as a spinon pairing 
(so-called 'spin gap').~\cite{rf:tanamoto} 
 The magnetic scenarios based on the anti-ferromagnetic or SDW gap formation 
or their precursor have been proposed by various authors.~\cite{rf:SDW} 

 Furthermore, the pairing scenarios as a precursor of the superconductivity 
are classified into several types. 
 The phase fluctuation scenarios have been proposed by 
Emery and Kivelson~\cite{rf:emery} and calculated by various 
authors.~\cite{rf:phase} 
 The scenario based on the strong coupling superconductivity has been 
proposed~\cite{rf:randeriareview} on the basis of the famous 
Nozi$\grave{{\rm e}}$res and Schmitt-Rink formalism.~\cite{rf:Nozieres} 
 The Nozi$\grave{{\rm e}}$res and Schmitt-Rink formalism is justified in the 
low density limit. However, the nearly half-filled lattice system 
should be regarded as a rather high density case. 
 Therefore, the Nozi$\grave{{\rm e}}$res and Schmitt-Rink formalism cannot be 
applied to the pseudogap phenomena in High-$T_{{\rm c}}$ cuprates. 
 Our scenario is based on the strong coupling superconductivity, but is 
different from the Nozi$\grave{{\rm e}}$res and Schmitt-Rink 
formalism. 
 We think of the pseudogap as the gap brought about by the resonance 
scattering~\cite{rf:resonance,rf:janko} with the strong superconducting 
fluctuations. The strong superconducting fluctuations necessarily exist in 
case of the strong coupling superconductivity in the quasi-two dimensional 
systems. We have shown that the pseudogap phenomena are naturally understood 
on the basis of the resonance scattering scenario.~\cite{rf:yanasepg} 
 
 Recently, the magnetic field effects on the NMR spin-lattice relaxation 
rate $1/T_{1}$ have been measured and discussed by several groups 
to determine the correct scenarios for the pseudogap 
phenomena~\cite{rf:zheng,rf:gorny,rf:mitrovic}. 
 The experimental results are interpreted as follows.  
The magnetic field effects cannot be observed in under-doped cuprates in which 
the strong pseudogap phenomena occurs in the wide temperature 
region.~\cite{rf:zheng,rf:gorny} 
 In particular, the onset temperature $T^{*}$ does not vary. 
 On the other hand, the magnetic field effects are visible 
from optimally-doped to slightly over-doped cuprates in which only the weak 
pseudogap phenomena are observed in the narrow temperature 
region.~\cite{rf:mitrovic,rf:zhengprivate} The observed magnetic field 
dependences are explained by the conventional weak coupling 
theory.~\cite{rf:eschrig,rf:zhengprivate} 
 However, for the under-doped cuprates, we have no theoretical explanation of 
the magnetic field effects on the pseudogap phenomena. 

 In this paper, we point out that the magnetic field effects are naturally and 
continuously understood from under-doped to over-doped cuprates 
on the basis of our resonance scattering scenario. 
 In particular, there is an interpretation that regards the experimental 
results for under-doped systems as a negative evidence for the pairing 
scenario.~\cite{rf:zheng}  
 Our results conclude that this interpretation is inappropriate. 
 It is generally considered that the superconducting fluctuations are 
remarkably influenced by the magnetic field, while the effects of the 
magnetic field on the spin-fluctuations are considered to be small. 
 Therefore, the experimental results may be interpreted as an evidence for 
the magnetic scenario for the pseudogap. 
 The misinterpretation is caused by the loss of the understanding for  
the strong coupling superconductivity. 
 Therefore, we give an explanation for the magnetic field effects on the 
pseudogap phenomena on the basis of the strong coupling superconductivity. 
 Actually, the experimental results including their doping dependence rather 
support our scenario for the pseudogap phenomena.~\cite{rf:yanasepg}  
 
 This paper is constructed as follows. 
 In \S2, we give a model Hamiltonian and explain the theoretical framework  
adopted in this paper. In \S3, we explicitly calculate the single particle 
self-energy $ {\mit{\it \Sigma}}^{{\rm R}} (\mbox{\boldmath$k$}, \omega) $, 
density of states $\rho(\varepsilon)$, 
NMR spin-lattice relaxation rate $1/T_{1}T$ and their magnetic 
field dependences. In \S4, we discuss the transport phenomena in the pseudogap 
phase. In \S5, we summarize the obtained results and give discussions.

\section{Theoretical Framework}

 In this section, we describe the theoretical framework in this paper. 
 We calculate the magnetic field effects on the pseudogap phenomena by using 
the same formalism as is used in our previous paper.~\cite{rf:yanasepg}
 Therefore, in the first subsection \S2.1, we briefly explain the formalism 
and show the outline of the obtained results in ref.8. 
 In the second subsection \S2.2, we introduce the magnetic field effects 
thorough the Landau quantization and give a rough estimate for the effects. 
 Hereafter, we adopt the unit $\hbar=c=k_{{\rm B}}=1$. 

\subsection{Pseudogap phenomena under zero magnetic field} 

 We adopt the following two-dimensional model Hamiltonian which has a 
$d_{x^2-y^2}$-wave superconducting ground state, 
with High-$T_{{\rm c}}$ cuprates in mind.

\begin{eqnarray}
  \label{eq:model}
  H =  \sum_{\mbox{\boldmath$k$},s}  \varepsilon_{\mbox{\boldmath$k$}} 
c_{\mbox{\boldmath$k$},s}^{\dag} c_{\mbox{\boldmath$k$},s}  
 +  \sum_{\mbox{\boldmath$k$},\mbox{\boldmath$k'$},\mbox{\boldmath$q$}} 
V_{\mbox{\boldmath$k$}-\mbox{\boldmath$q$}/2,
\mbox{\boldmath$k'$}-\mbox{\boldmath$q$}/2} 
c_{\mbox{\boldmath$q$}-\mbox{\boldmath$k'$},\downarrow}^{\dag} 
c_{\mbox{\boldmath$k'$},\uparrow}^{\dag} 
c_{\mbox{\boldmath$k$},\uparrow} 
c_{\mbox{\boldmath$q$}-\mbox{\boldmath$k$},\downarrow}, 
\end{eqnarray}

 where $ V_{\mbox{\boldmath$k$},\mbox{\boldmath$k'$}} $ is the 
$d_{x^2-y^2}$-wave separable pairing interaction, 

\begin{eqnarray}
  \label{eq:d-wave}
  V_{\mbox{\boldmath$k$},\mbox{\boldmath$k'$}} = 
g \varphi_{\mbox{\boldmath$k$}} \varphi_{\mbox{\boldmath$k'$}}, \\
  \varphi_{\mbox{\boldmath$k$}} = \cos k_{x}-\cos k_{y}.
\end{eqnarray}

 Here, $g$ is negative. $ \varphi_{\mbox{\boldmath$k$}} $ is the 
$d_{x^2-y^2}$-wave form factor. 

 We consider the dispersion $\varepsilon_{\mbox{\boldmath$k$}}$ given by the 
tight-binding model for a square lattice including the nearest- and 
next-nearest-neighbor hopping $t$, $t'$, respectively, 

\begin{eqnarray}
 \label{eq:dispersion}
    \varepsilon_{\mbox{\boldmath$k$}} = -2 t (\cos k_{x} +\cos k_{y}) + 
                  4 t' \cos k_{x} \cos k_{y} - \mu. 
\end{eqnarray}

 We fix the lattice constant $ a = 1$. 
 We adopt $t=0.5 \rm{eV}$ and $t'=0.45 t$. These parameters well reproduce 
the Fermi surface of the typical High-$T_{{\rm c}}$ cuprates,
${\rm Y}{\rm Ba}_{2}{\rm Cu}_{3}{\rm O}_{6+\delta}$ and 
${\rm Bi}_{2}{\rm Sr}_{2}{\rm Ca}{\rm Cu}_{2}{\rm O}_{8+\delta}$. 
 We choose the chemical potential $\mu$ so that the filling $n=0.9$. 
This filling corresponds to the hole doping $\delta=0.1$. 
 The Fermi surface is shown in Fig.1. 

 \begin{figure}[ht]
   \begin{center}
   \epsfysize=4cm
    $$\epsffile{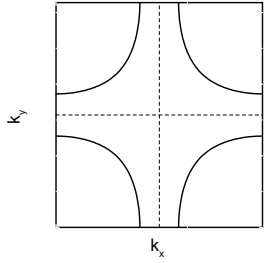}$$
     \caption{The Fermi surface adopted in this paper.}
     \label{fig:Fermi-surface}
   \end{center}
 \end{figure}

 In reality, the origin of the pairing interaction should be considered to be 
the anti-ferromagnetic spin fluctuations.~\cite{rf:monthoux,rf:moriya} 
 The spin fluctuations not only cause the pairing interaction but also affect 
the electronic state.~\cite{rf:monthoux,rf:moriya,rf:stojkovic,rf:yanasetr} 
 There are studies dealing with the pairing correlation arising from 
the spin fluctuations on the basis of the fluctuation exchange (FLEX) 
approximation.~\cite{rf:dahm,rf:koikegamisuper} 
 However, we do not introduce these effects 
because these details do not seriously affect the pseudogap phenomena 
as a precursor of the $d_{x^{2}-y^{2}}$-wave superconductivity. 
 There is a feedback effect on the pairing interaction 
arising from the pseudogap. 
 The pseudogap affects the low frequency component of the spin 
fluctuations. However, the pairing interaction is mainly caused by 
the high frequency component of the spin fluctuations. 
 Therefore, we can neglect the feedback effect on the pairing interaction  
and start from the model with an attractive interaction. 

 The superconducting fluctuations are expressed by the T-matrix (Fig.2), 

\begin{eqnarray}
  \label{eq:t-matrix}
  t(\mbox{\boldmath$q$},{\rm i} \Omega_{n})^{-1} & = & 
  g^{-1} + \chi_{0}(\mbox{\boldmath$q$},{\rm i} \Omega_{n}), 
\\
  \chi_{0}(\mbox{\boldmath$q$},{\rm i} \Omega_{n}) &  = &
  T \sum_{\mbox{\boldmath$k'$},\omega_{m}} 
   {\mit{\it G}} (\mbox{\boldmath$k'$},{\rm i} \omega_{m}) 
   {\mit{\it G}} (\mbox{\boldmath$q$}-\mbox{\boldmath$k'$},
   {\rm i} \Omega_{n} - {\rm i} \omega_{m})
   \varphi_{\mbox{\boldmath$k'$}-\mbox{\boldmath$q$}/2}^{2}.
\end{eqnarray}

 Here, $\omega_{m} = 2 \pi (m+\frac{1}{2}) T$ and $\Omega_{n}= 2 \pi n T$ are 
the fermionic and bosonic Matsubara frequencies, respectively.

\begin{figure}[ht]
  \begin{center}
   \epsfysize=5cm
$$\epsffile{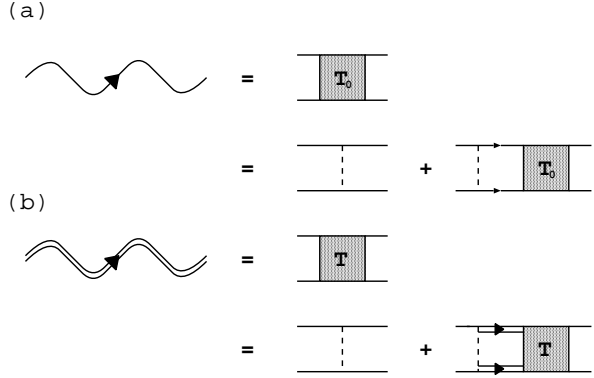}$$
    \caption{The scattering vertex represented by the ladder diagrams 
             in the particle-particle channel (T-matrix).  
             The dashed lines represent the attractive interaction. 
             The single and double solid lines represent the propagators of 
             the bare and renormalized fermions, respectively. 
             The single and double wavy lines represent the propagators of 
             the bare and renormalized fluctuating Cooper pairs, 
             respectively. 
             }             
    \label{fig:scattering-vertex}
  \end{center}
\end{figure}

 Here, the scattering vertex arising from the superconducting fluctuations, 
$\Gamma(\mbox{\boldmath$k$},\mbox{\boldmath$q$}-\mbox{\boldmath$k$}:
\mbox{\boldmath$k'$},\mbox{\boldmath$q$}-\mbox{\boldmath$k'$}:
{\rm i} \Omega_{n})$ is 
factorized into 
$\Gamma(\mbox{\boldmath$k$},\mbox{\boldmath$q$}-\mbox{\boldmath$k$}:
\mbox{\boldmath$k'$},\mbox{\boldmath$q$}-\mbox{\boldmath$k'$}:
{\rm i} \Omega_{n}) = 
\varphi_{\mbox{\boldmath$k$}-\mbox{\boldmath$q$}/2} 
t(\mbox{\boldmath$q$},{\rm i} \Omega_{n}) 
\varphi_{\mbox{\boldmath$k'$}-\mbox{\boldmath$q$}/2}$. 
 The form factor $\varphi_{\mbox{\boldmath$k$}}$ in the above expression 
gives rise to the $d_{x^{2}-y^{2}}$-wave shape of the pseudogap.

 When  $ 1 + g \chi_{0}(\mbox{\boldmath$0$},0) = 0 $, 
$t(\mbox{\boldmath$0$},0)$ diverges and the superconductivity occurs. 
 This is the famous Thouless criterion which is equivalent to that of 
the BCS theory in the weak coupling limit.~\cite{rf:Nozieres} 
 Analytically continued T-matrix $ t(\mbox{\boldmath$q$},\Omega) $ can be 
regarded as a propagator of the fluctuating Cooper pairs. 

 Here, we are interested in the normal state near the superconducting 
critical point, where the superconducting fluctuations are enhanced. 
 There, $ 1 + g \chi_{0}(\mbox{\boldmath$0$},0) $ is small and 
$ t(\mbox{\boldmath$q$},\Omega) $ is strongly enhanced around 
$ \mbox{\boldmath$q$} = \Omega = 0 $. 
 Even in the weak coupling limit, there are corrections on the various 
quantities due to the superconducting fluctuations. They are well known as  
the Aslamazov-Larkin term (AL term)~\cite{rf:AL} and the Maki-Thompson term 
(MT term).~\cite{rf:MT} 
 These terms are the corrections on the two-body correlation function. 
 On the other hand, the superconducting fluctuations more seriously affect 
the one-particle electronic states in the strong or intermediate coupling 
region. The superconducting fluctuations give rise to the pseudogap 
phenomena. 
 The weak correction on the density of states (DOS correction term) has been 
discussed for High-$T_{{\rm c}}$ cuprates within the weak coupling 
theory.~\cite{rf:eschrig,rf:DOS} 
 Our calculation corresponds to an extension of these weak coupling 
theories to the strong coupling ones.
 
 Because $ t(\mbox{\boldmath$q$},\Omega) $ is strongly enhanced around 
$ \mbox{\boldmath$q$} = \Omega = 0 $, 
its main contribution to the single particle self-energy 
$ {\mit{\it \Sigma}}(\mbox{\boldmath$k$},\omega) $ originates from 
the vicinity of $ \mbox{\boldmath$q$} = \Omega = 0 $.
 Therefore, we expand $ t^{-1}(\mbox{\boldmath$q$},\Omega) $ 
in the vicinity of $ \mbox{\boldmath$q$} = \Omega = 0 $. 
 This expansion corresponds to the time-dependent-Ginzburg-Landau (TDGL) 
expansion.

\begin{eqnarray}
  \label{eq:TDGL}
   g t^{-1}(\mbox{\boldmath$q$},\Omega) = 
   t_{0} + b \mbox{\boldmath$q$}^{2} - (a_{1}+{\rm i}a_{2}) \Omega. 
\end{eqnarray}

 The properties of the TDGL parameters are discussed in detail in our previous 
paper.~\cite{rf:yanasepg} The outline is the following. 
 As is described above, $ t_{0} = 1 + g \chi_{0}(\mbox{\boldmath$0$},0) $ is 
$0$ at the critical point and is sufficiently small in the vicinity of the 
critical point. 
 The parameter $ b $ is generally related to the coherence length $\xi_{0}$, 
$b \propto \xi_{0}^{2}$. 
 The parameter $a_{2}$ express the time scale of the fluctuations. 
 Roughly speaking, the parameters $a_{2}$ and $b$ are described as 
 
\begin{eqnarray}
 a_{2} \propto \rho_{{\rm d}}(0)/T_{{\rm c}} 
\nonumber \\
 b \propto \rho_{{\rm d}}(0)/T_{{\rm c}}^{2}. 
\end{eqnarray}

 Here, we have defined the effective density of states 
for the $ d_{x^{2}-y^{2}} $-wave symmetry, $ \rho_{{\rm d}}(\varepsilon) 
=  \sum_{\mbox{\boldmath$k$}} \rho_{\mbox{\boldmath$k$}}(\varepsilon) 
  \varphi_{\mbox{\boldmath$k$}}^{2} $, 
 where, $ \rho_{\mbox{\boldmath$k$}}(\varepsilon) $ is the one-particle 
spectral weight $ \rho_{\mbox{\boldmath$k$}}(\varepsilon) 
=  A_{\mbox{\boldmath$k$}}(\varepsilon) = 
- \frac{1}{\pi} {\rm Im} {\mit{\it G}}^{{\rm R}} 
(\mbox{\boldmath$k$},\varepsilon) $. 
 It should be noticed that $\rho_{{\rm d}}(\varepsilon)$ is more sensitive to 
the pseudogap formation rather than the usual density of states 
$  \rho(\varepsilon) = 
  \sum_{\mbox{\boldmath$k$}} \rho_{\mbox{\boldmath$k$}}(\varepsilon) $.

 Because of the high critical temperature $T_{{\rm c}}$ and the 
renormalization effect by the pseudogap, both $a_{2}$ and $b$ are strongly 
reduced in the strong coupling superconductivity. 
 These features indicate that the scattering vertex due to the superconducting 
fluctuations is strongly enhanced. 
 Although the T-matrix calculation 
used in this paper does not include the renormalization effect, these 
behaviors are obtained qualitatively. 
 
 On the other hand, $ a_{1} $ is not so reduced by the strong coupling 
superconductivity. 
 Especially, High-$T_{{\rm c}}$ cuprates have a comparatively large value of 
$ a_{1} $ because of their strong particle-hole asymmetry. 
 Therefore, we cannot neglect $ a_{1} $, although it is 
usually neglected in the weak coupling theories.~\cite{rf:ebisawa}  

 In the T-matrix calculation, we estimate the TDGL parameters by using 
the non-interacting Green function 
${\mit{\it G}}^{(0) {\rm R}} (\mbox{\boldmath$k$},\omega) = 
(\omega - \varepsilon_{\mbox{\boldmath$k$}} + {\rm i} \delta)^{-1} $ 
(Fig.2(a)). 
 This estimation corresponds to the Gauss approximation for the superconducting fluctuations. The critical temperature is determined in the mean field level, 
$T_{{\rm c}} = T_{{\rm MF}}$. 
 As will be discussed in \S5, the self-consistent T-matrix calculation 
includes the somewhat critical fluctuations (Figs.2(b) and 11). 
 However, fundamental features do not change.

 As is minutely described in the previous paper~\cite{rf:yanasepg}, 
the resonance scattering by the strong superconducting fluctuations gives 
rise to the pseudogap on both the one-particle spectral weight and the 
density of states. 

 In this paper we describe these features by using the T-matrix calculation 
(Fig.3(a)). Although we pointed out the several important points in the 
self-consistent T-matrix calculation (Fig.3(b)), 
the pseudogap is properly described within the T-matrix 
calculation.~\cite{rf:yanasepg}

\begin{figure}[ht]
  \begin{center}
   \epsfysize=6cm
$$\epsffile{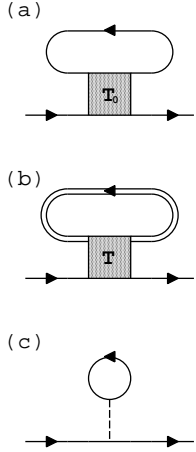}$$
    \caption{The diagrams of the single particle self-energy based on 
             (a) the T-matrix approximation calculated in this paper and 
             (b) the self-consistent T-matrix approximation, respectively.
             (c) The Hartree-Fock term which we exclude. 
              We consider that this term is included 
              in the dispersion relation 
              $\varepsilon_{\mbox{\boldmath$k$}}$ from the beginning.   }
    \label{fig:selfenergy-diagram}
  \end{center}
\end{figure}

 In the T-matrix calculation, the self-energy is given by 

\begin{eqnarray}
  \label{eq:selfenergymatubara}
  {\mit{\it \Sigma}} (\mbox{\boldmath$k$}, {\rm i} \omega_{n})  = 
  T \sum_{\mbox{\boldmath$q$},{\rm i} \Omega_{m}}
  t(\mbox{\boldmath$q$},{\rm i} \Omega_{m}) 
  {\mit{\it G}}^{(0)} (\mbox{\boldmath$q$}-\mbox{\boldmath$k$}, 
   {\rm i} \Omega_{m} - {\rm i} \omega_{n})
  \varphi_{\mbox{\boldmath$k$}-\mbox{\boldmath$q$}/2}^{2}.    
\end{eqnarray}

 After the analytic continuation, we obtain

\begin{eqnarray}
  \label{eq:selfenergyreal}
  {\mit{\it \Sigma}}^{{\rm R}} (\mbox{\boldmath$k$}, \omega) & = &
  \sum_{\mbox{\boldmath$q$}} \int \frac{{\rm d} \Omega}{\pi} 
  [ b(\Omega) {\rm Im} t(\mbox{\boldmath$q$},\Omega) 
   {\mit{\it G}}^{(0) {\rm A}} (\mbox{\boldmath$q$}-\mbox{\boldmath$k$}, 
    \Omega-\omega)
\nonumber \\
 & &  - f(\Omega) t(\mbox{\boldmath$q$},\Omega+\omega) 
   {\rm Im} {\mit{\it G}}^{(0) {\rm R}} 
   (\mbox{\boldmath$q$}-\mbox{\boldmath$k$},\Omega) ] 
   \varphi_{\mbox{\boldmath$k$}-\mbox{\boldmath$q$}/2}^{2}, 
\\ 
   {\rm Im} {\mit{\it \Sigma}}^{{\rm R}} (\mbox{\boldmath$k$}, \omega) & = & 
   -\sum_{\mbox{\boldmath$q$}} \int \frac{{\rm d} \Omega}{\pi} 
   [ b(\Omega+\omega)+f(\Omega) ] 
   {\rm Im} t(\mbox{\boldmath$q$},\Omega+\omega) 
   {\rm Im} {\mit{\it G}}^{(0) {\rm R}} 
    (\mbox{\boldmath$q$}-\mbox{\boldmath$k$},\Omega) 
   \varphi_{\mbox{\boldmath$k$}-\mbox{\boldmath$q$}/2}^{2}. 
\end{eqnarray}

 Here,$f(\Omega)$ and $b(\Omega)$ are the Fermi and Bose distribution 
functions, respectively. 

 We explicitly estimate the TDGL expansion parameters and numerically 
calculate the single particle self-energy 
$ {\mit{\it \Sigma}}^{{\rm R}} (\mbox{\boldmath$k$}, \omega) $. 
 Typical features of the single-particle self-energy are shown 
in Fig.4. Here, we exclude the trivial Hartree-Fock term shown in Fig.3(c). 
 It is notable that the real part of the self-energy has a positive slope 
at the low frequency, and the imaginary part has a sharp peak 
in its absolute value there. 
 Both features are anomalous compared with the conventional Fermi liquid 
theory. These anomalous features of the single particle self-energy should be 
regarded as the effects of the resonance scattering. 
 Of course, the resonance scattering becomes strong as the superconductivity 
becomes strong coupling one and the systems approach the critical point. 

 \begin{figure}[ht]
   \begin{center}
   \epsfysize=6cm
$$\epsffile{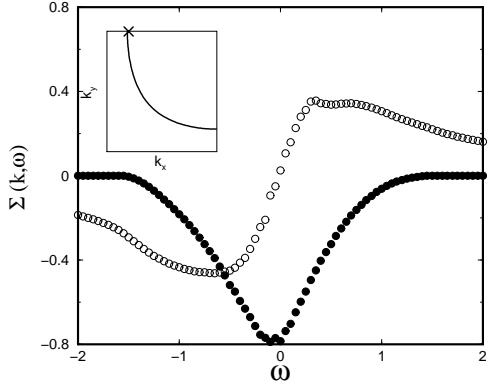}$$
    \caption{The single particle self-energy 
              $ {\mit{\it \Sigma}}^{{\rm R}} (\mbox{\boldmath$k$}, \omega) $ 
              on the Fermi surface 
              near $(0,\pi)$. The open and closed circles show the real part 
              and the imaginary part, respectively.
              Here, $\mbox{\boldmath$k$}=(\pi/5,\pi)$, $g=-1.0$, $T=0.21$ and 
              $4 {\rm e} B=0.01$. 
              The $\mbox{\boldmath$k$}$-point is shown in the inset. 
              Here, $T_{{\rm c0}}=T_{{\rm MF}}=0.196$. $T_{{\rm c0}}$ is 
              $T_{{\rm c}}$ at $B=0$. }
     \label{fig:T-matrix-selfenergy}
   \end{center}
 \end{figure}

 These features lead to the pseudogap. The corresponding one-particle 
spectral weight $ A(\mbox{\boldmath$k$}, \omega) $ and density of states 
$ \rho(\omega) $ are shown in Fig.5. 
 Both show the gap structure above $T_{{\rm c}}$. 
 It should be noticed that the pseudogap is the characteristics of the strong 
coupling superconductivity and does not occur in the weak coupling limit. 
 In Fig.4 and 5, we have included the magnetic field effect described 
in the next subsection.

 \begin{figure}[ht]
   \begin{center}
   \epsfysize=6cm
$$\epsffile{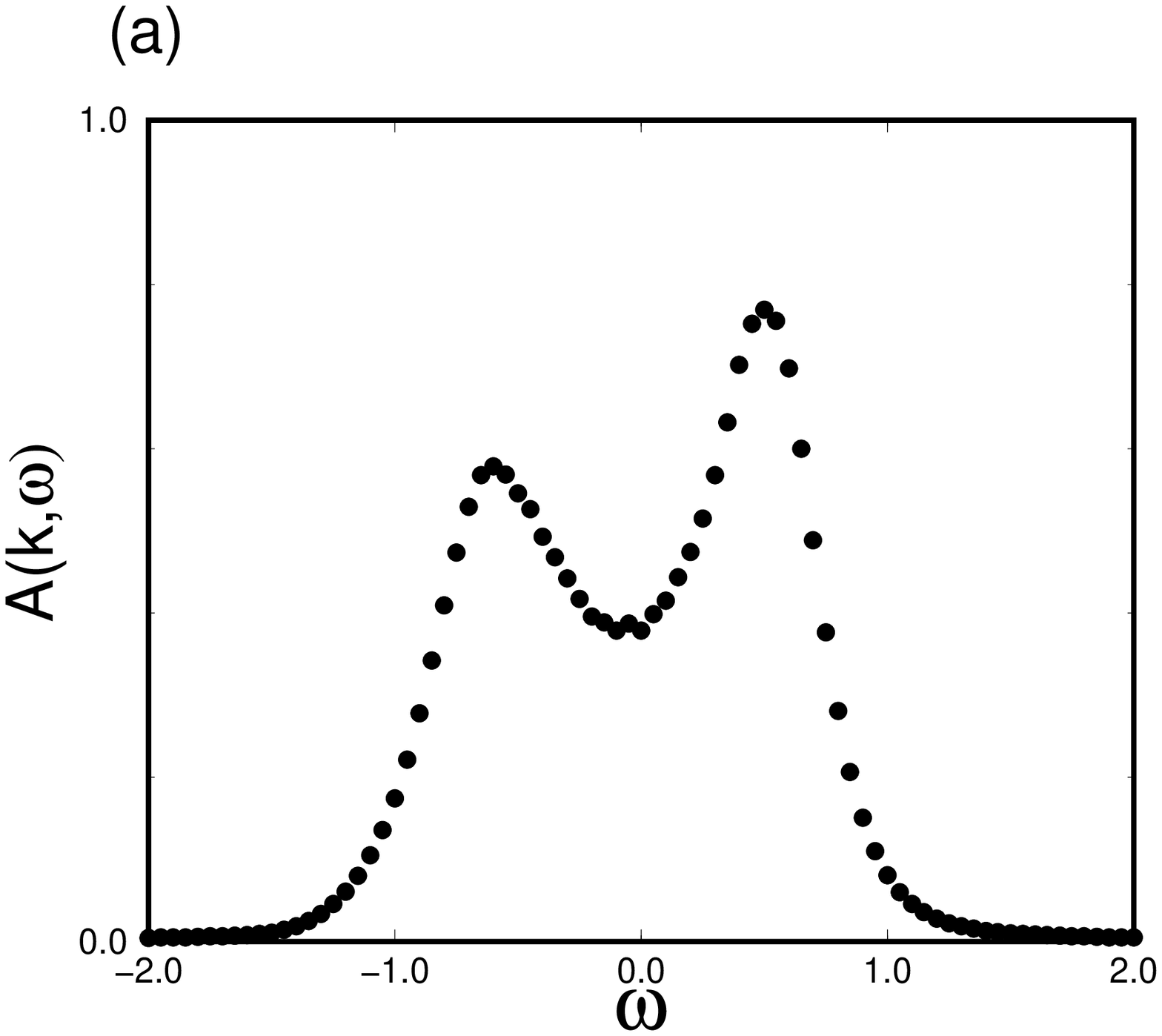}$$
   \epsfysize=6cm
$$\epsffile{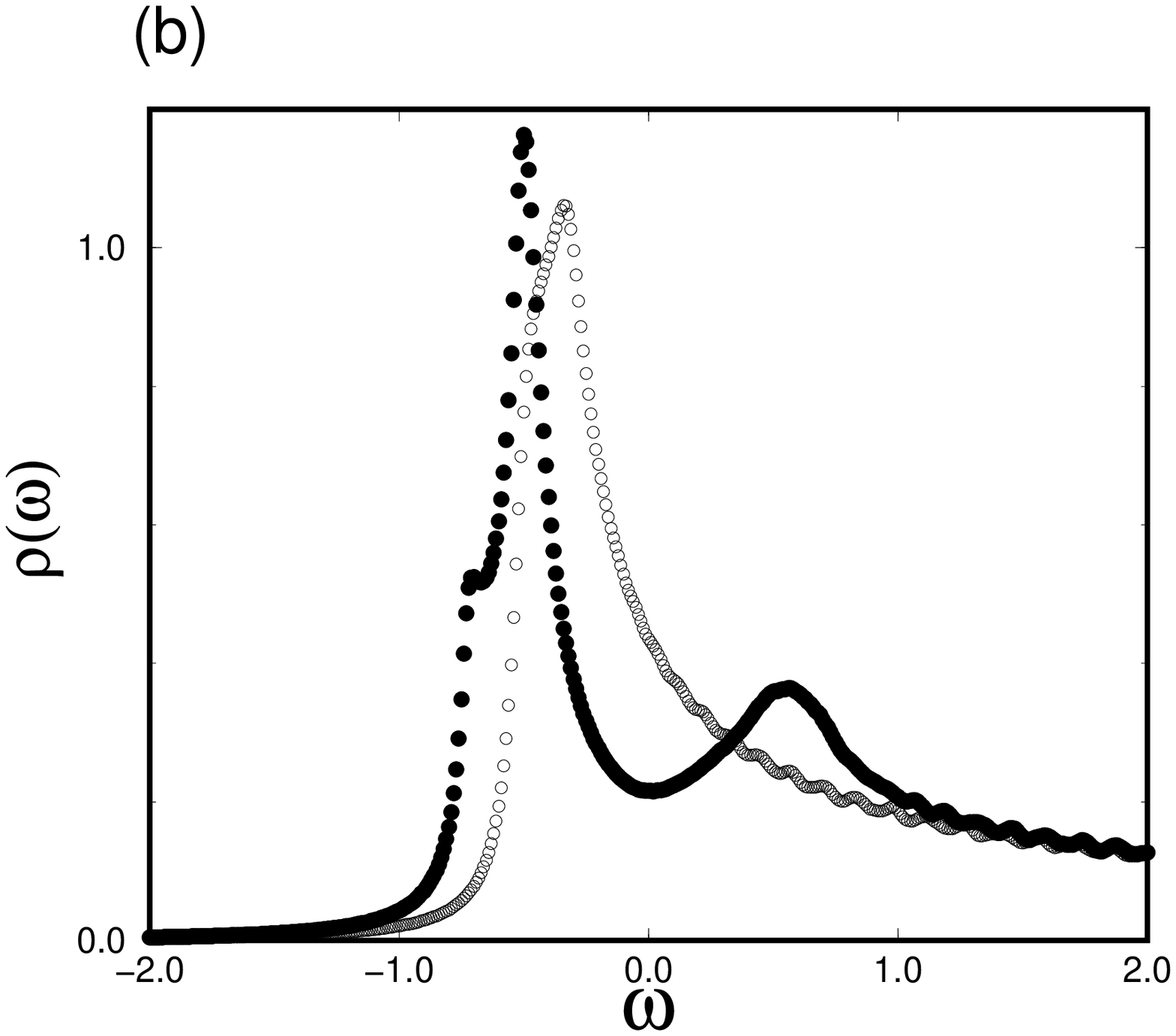}$$
    \caption{ (a) The one-particle spectral weight 
                  $ A(\mbox{\boldmath$k$}, \omega) $ at 
                  $\mbox{\boldmath$k$}=(\pi/5,\pi)$.
              (b) The density of states $ \rho(\omega) $. 
                  The open circles show the non-interacting density of states 
                  in our model, and the closed circles show the calculated 
                  result. 
                  The other parameters are the same as those in Fig.4.   }
     \label{fig:T-matrix-temperature}
   \end{center}
 \end{figure}

 The pseudogap reduces the critical temperature $T_{{\rm c}}$. 
 The reduction becomes more remarkable as the coupling constant increases. 
 Therefore, although the mean field critical temperature $T_{{\rm MF}}$ 
remarkably increases with the coupling constant $|g|$, $T_{{\rm c}}$
does not vary so much.~\cite{rf:yanasepg}

 Here, $T_{{\rm c}}$ is scaled by the effective Fermi energy 
$\varepsilon_{{\rm F}}$. 
 By considering the fact, it is naturally understood that 
$T_{{\rm c}}$ decreases with the doping quantity in the under-doped region. 
 As the doping quantity decreases, the system approaches to 
the Mott insulator. Therefore, it should be considered that 
the renormalization effect for the effective Fermi energy 
$\varepsilon_{{\rm F}}$ is enhanced with decreasing the doping quantity. 
 Since $T_{{\rm c}}/\varepsilon_{{\rm F}}$ is almost independent of 
the coupling constant $|g|$ in the strong coupling region, 
$T_{{\rm c}}$ decreases with $\varepsilon_{{\rm F}}$ in the under-doped 
region. 
 Thus, our theory naturally and appropriately explains 
the pseudogap phenomena in High-$T_{{\rm c}}$ cuprates.

\subsection{Magnetic field effects on the pseudogap phenomena}

 In this subsection, we introduce the magnetic field effect. 
 In this paper, we consider the magnetic field applied along the {\it c}-axis, 
$B \parallel \vec{c}$. 
 The main effect of the magnetic fields is the Landau level quantization for 
the superconducting fluctuations. It corresponds to the quantization of the 
orbital motion of the fluctuating Cooper pairs. 
 The quantization is expressed by the replacement of the quadratic term 
of the momentum as 
$\mbox{\boldmath$q$}^{2} \Rightarrow 4{\rm e} B (n+\frac{1}{2}) 
$.~\cite{rf:eschrig} 

 The Landau quantization has the following two important effects. 
One is the Landau degeneracy which generally enhances the fluctuations. 
The Landau degeneracy reduces the dimensionality of the fluctuations. 
 The other is the suppression of the superconductivity which weakens 
the pseudogap. The distance to the critical point increases 
as $t_{0} \Rightarrow t_{0} + 2 b {\rm e} B $. This corresponds to the energy 
level of the Lowest Landau level.  
 When considering at the fixed temperature, the dominant effect is the latter. 
 We can see that the characteristic magnetic field 
$ B_{{\rm ch}} $ for the pseudogap phenomena is scaled by the quantity 
$t_{0}/b$, that is $ B_{{\rm ch}} \propto t_{0}/b $.~\cite{rf:yanasepro} 
 The ratio $b/t_{0}$ corresponds to the square of the GL correlation length 
$\xi_{{\rm GL}}$ 
for the superconducting fluctuations, that is, $b/t_{0} = \xi_{{\rm GL}}^{2}$. 
 The magnetic field effects are scaled by the quantity $B \xi_{{\rm GL}}^{2}$. 
 Thus, the pseudogap is affected by the magnetic field according to 
the magnetic flux penetrating the correlated area $\xi_{{\rm GL}}^{2}$. 

 As we mentioned above, the parameter $b$ is small in the strong coupling case.
 Moreover, the fact that the pseudogap phenomena take place in the wide 
temperature region means that $t_{0}$ is large near the pseudogap onset 
temperature $T^{*}$. 
 As a result, the characteristic magnetic field $ B_{{\rm ch}} $ is large, 
especially near $T^{*}$. In other words, the magnetic 
field effects are remarkably small in case of the strong 
coupling superconductivity. Especially, the onset temperature $T^{*}$ does not 
vary. Since $\xi_{{\rm GL}}$ diverges at the critical temperature 
$T_{{\rm c}}$, the magnetic field effects are sure to appear near 
$T_{{\rm c}}$. 
However, the region in which the effects appear is remarkably small. 
 On the other hand, in the relatively weak coupling case, the magnetic field 
dependence is large and $T^{*}$ may vary. 
 These features well explain the results of the high field NMR measurements
including their doping 
dependence.~\cite{rf:zheng,rf:gorny,rf:mitrovic,rf:zhengprivate}
 
 Here, we have neglected the Zeeman coupling term. 
Although the Zeeman coupling term plays an important role at the low 
temperature in superconducting state,~\cite{rf:FFLO} 
it has only higher order correction in the fluctuating region. 
 This fact can be simply understood as follows. The lowest order correction of 
the Zeeman coupling term on the superconducting fluctuations is the second 
order and described as $ 4{\rm e} B (n+\frac{1}{2}) \Rightarrow 
4{\rm e} B (n+\frac{1}{2}) + \frac{8}{v^{2}} (\mu B)^{2} $. 
 Here, $v$ is a mean value of the quasi-particle velocity on the Fermi surface.
 $\mu = g \mu_{{\rm B}}/2$ is the magnetic moment of the electrons. Here, 
$g$ is the $g$-value and $\mu_{{\rm B}}$ is the Bohr magneton. 
 Thus, the Zeeman coupling term slightly weaken the superconducting 
fluctuations. 
 However, it has only higher order effect with respect to the magnetic field 
compared with the Landau quantization. 
Therefore, the effect of the Zeeman coupling term is 
extraordinary small in the weak coupling limit since the typical magnetic 
field is small. 
 Also for High-$T_{{\rm c}}$ cuprates, it is higher order and remarkably small 
compared with the effect of the Landau quantization in the magnetic field 
of the experimentally relevant order. 
 Actually, the magnetic field adopted in this paper is the order of 
$ {\rm e} B \sim 10^{-2} $ in our unit. That corresponds to 
$B \sim 10 {\rm Tesla}$. In this case, the effect of the Zeeman coupling term 
is higher order than that of Landau quantization as $10^{-2}$. 
 Thus, it is justified to neglect the Zeeman coupling term. 
 Of course, we cannot neglect the Zeeman coupling term under the extraordinary 
high magnetic field in the strong coupling limit. However, such an extreme 
situation is not realistic. When the magnetic field is applied 
perpendicular to the {\it c}-axis $B \perp \vec{c} $, 
effect of the Zeeman coupling term is relatively important because the 
coherence length along the {\it c}-axis $\xi_{{\rm c}}$ is small.

\section{Magnetic Field Dependence of the NMR Spin-Lattice Relaxation Rate, 
$1/T_{1}T$}

 In this section, we actually calculate the NMR spin-lattice relaxation rate 
$1/T_{1}T$ under the magnetic field with the recent high field NMR 
measurements in mind. 
 We calculate $1/T_{1}T$ by using the general expression, 

\begin{eqnarray}
  1/T_{1}T = \sum_{\mbox{\boldmath$q$}} |A(\mbox{\boldmath$q$})|^{2}  
             [\frac{1}{\omega} 
             {\rm Im} \chi_{{\rm s}}^{{\rm R}} (\mbox{\boldmath$q$}, \omega) 
             \mid_{\omega \to 0}].
\end{eqnarray}

 Here, we neglect the momentum dependence of the hyperfine coupling 
$ A(\mbox{\boldmath$q$}) $ for simplicity, which does not affect the 
magnetic field dependence of $1/T_{1}T$. 

 We calculate the spin susceptibility 
$ \chi_{{\rm s}}^{{\rm R}} (\mbox{\boldmath$q$}, \omega) $ which corresponds 
to the two-body correlation function shown in Fig.6.

\begin{figure}[ht]
  \begin{center}
   \epsfysize=2.5cm
$$\epsffile{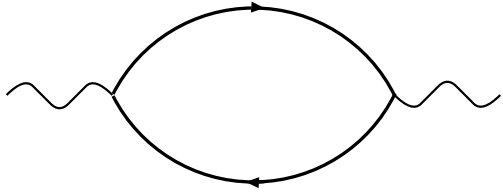}$$
    \caption{The diagram representing the dynamical spin susceptibility 
             $ \chi_{{\rm s}}^{{\rm R}} (\mbox{\boldmath$q$}, \omega) $} 
    \label{fig:1/T1T-diagram}
  \end{center}
\end{figure}

 Here, the solid lines are the renormalized Green function 
${\mit{\it G}}^{{\rm R}} (\mbox{\boldmath$k$},\omega) = 
(\omega - \varepsilon_{\mbox{\boldmath$k$}} - 
{\mit{\it \Sigma}}^{{\rm R}} (\mbox{\boldmath$k$},\omega))^{-1} $. 
 The self-energy ${\mit{\it \Sigma}}^{{\rm R}} (\mbox{\boldmath$k$},\omega) $
is calculated by using the T-matrix approximation as we described before. 
 The effects of the superconducting fluctuations are included in the 
self-energy. 

 In calculating the self-energy ${\mit{\it \Sigma}}^{{\rm R}} 
(\mbox{\boldmath$k$},\omega) $, we linearize the dispersion relation as 
$ \varepsilon_{\mbox{\boldmath$k$}-\mbox{\boldmath$q$}} = 
\varepsilon_{\mbox{\boldmath$k$}} - \mbox{\boldmath$v$}_{\mbox{\boldmath$k$}} 
\mbox{\boldmath$q$} $. This linearization is justified because the only small 
region around $\mbox{\boldmath$q$} = 0$ contributes to the self-energy. 
 We replace the quadratic term as 
$\mbox{\boldmath$q$}^{2} \Rightarrow 4{\rm e} B (n+\frac{1}{2})$. 
 This process corresponds to the Landau level quantization for the 
superconducting fluctuations. 
 
 From eq.(3.1), $1/T_{1}T$ is expressed as, 

\begin{eqnarray}
  1/T_{1}T & = & - \sum_{\mbox{\boldmath$k$},\mbox{\boldmath$q$}} \int 
             \frac{{\rm d}\omega}{\pi} f'(\omega) 
             {\rm Im} {\mit{\it G}}^{{\rm R}} (\mbox{\boldmath$k$},\omega) 
             {\rm Im} {\mit{\it G}}^{{\rm R}} (\mbox{\boldmath$k+q$},\omega) 
\nonumber \\
           & = & - \pi \int {\rm d}\omega f'(\omega) \rho(\omega)^{2}.
\end{eqnarray}

 Here, $f'(\omega)$ is the differential of the Fermi distribution function. 
 This expression is reduced to the well-known expression 
$ 1/T_{1}T = \pi \rho(0)^{2} $ at $ T=0 $. 
 After all, we calculate the decrease of $ 1/T_{1}T $ by the suppression of 
the density of states. 

 Generally speaking, we can consider the Aslamazov-Larkin term (AL term) and 
the Maki-Thompson term (MT term) as corrections by the fluctuations on the 
two-body correlation function.~\cite{rf:AL,rf:MT} 
 However, the AL term dose not exist in calculating the spin susceptibility 
$ \chi_{{\rm s}}^{{\rm R}} (\mbox{\boldmath$q$}, \omega) $. 
 We can understand this fact by considering the spin index for the spin 
singlet pairing.~\cite{rf:DOS,rf:eschrig}  
 The contribution from the MT term is small in case of the d-wave pairing, and 
suppressed by the slight elastic scattering.~\cite{rf:eschrig}  
 Therefore, we have only to calculate the decrease of $ 1/T_{1}T $ by the 
pseudogap as the effect of the superconducting fluctuations. 

 Of course, we have to take account of the anti-ferromagnetic 
spin-fluctuations in order to describe the whole temperature dependence of 
$ 1/T_{1}T $.
 $ 1/T_{1}T $ increases owing to the anti-ferromagnetic spin fluctuations 
in the normal phase ($ T > T^{*} $), and decreases owing to the 
superconducting fluctuations in the pseudogap phase 
($ T^{*} > T > T_{{\rm c}} $). 
 Generally speaking, the magnetic field is considered to have a great effect 
on the superconducting fluctuations, while the effect on the 
spin-fluctuations is comparatively small. 
 Because we pay attention to the magnetic field dependence in this paper, 
we have only to calculate the decrease of 
$ 1/T_{1}T $ due to the superconducting fluctuations and its magnetic field 
dependence. 
 Actually, the misinterpretation for the experimental results is caused by the 
loss of the understanding of the magnetic field effect on the superconducting 
fluctuations in case of the strong coupling superconductivity. 
 Our calculation gives a clear understanding about the magnetic field 
dependence of the pseudogap phenomena. 

 Even if the effect of the exchange enhancement is taken into account, 
the results for the magnetic field effect do not change, qualitatively. 
 At the last of this section, we definitely calculate the effect of the 
exchange enhancement within the random phase approximation (RPA). 
 Qualitatively, the same results are given there.

 The calculated results are shown in Fig.7, 8 and 9. 
 In all figures, the magnetic field is varied as $ 4 {\rm e} B = 0.01, 0.02, 
0.05$ and $ 0.1 $ in our unit. The horizontal axis is the temperature scaled 
by the zero-field critical temperature $T_{{\rm c}0}$. 
 All results can be understood by considering the characteristic magnetic 
field, as we have mentioned in the previous section, 

\begin{eqnarray}
  B_{{\rm ch}} \propto t_{0}/b = \xi_{{\rm GL}}^{-2}. 
\end{eqnarray}

 The results for the relatively weak coupling case $ g = -0.5 $ are shown 
in Fig.7. In this case, only the weak pseudogap is observed in the narrow  
temperature region. We consider that this case corresponds to the slightly 
over-doped or optimally-doped cuprates.  
 The magnetic field dependence of $ 1/T_{1}T $ is clearly 
observed and $T^{*}$ varies. These behaviors are consistent with the NMR 
experiments in the slightly over-doped cuprates.~\cite{rf:zhengprivate}

\begin{figure}[ht]
  \begin{center}
   \epsfysize=6cm
$$\epsffile{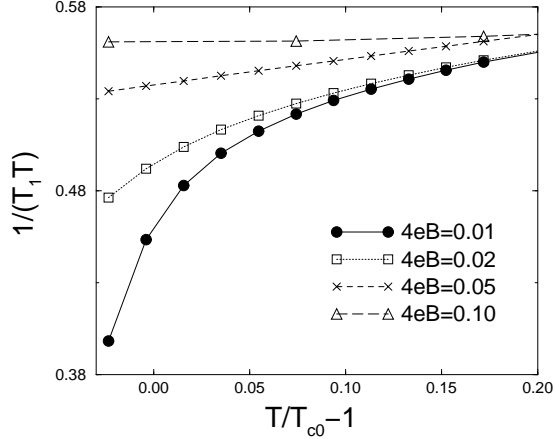}$$
     \caption{The calculated results for $1/T_{1}T$ under the various magnetic 
             field $ 4 {\rm e} B = 0.01, 0.02, 0.05, 0.1 $. 
             The weak pseudogap case $ g = -0.5 $.}
  \end{center}
\end{figure}

 In case of $ g = -0.8 $, the magnetic field dependence is small 
since the parameter $b$ decreases(Fig.8). 
 In particular, $ 1/T_{1}T $ is almost 
independent of the magnetic field near the onset temperature $T^{*}$ 
where the parameter $t_{0}$ is large. On the other hand, the magnetic field 
dependence can be observed in the vicinity of the critical temperature 
$T_{{\rm c}}$, since $t_{0}$ is small there.

\begin{figure}[ht]
  \begin{center}
   \epsfysize=6cm
$$\epsffile{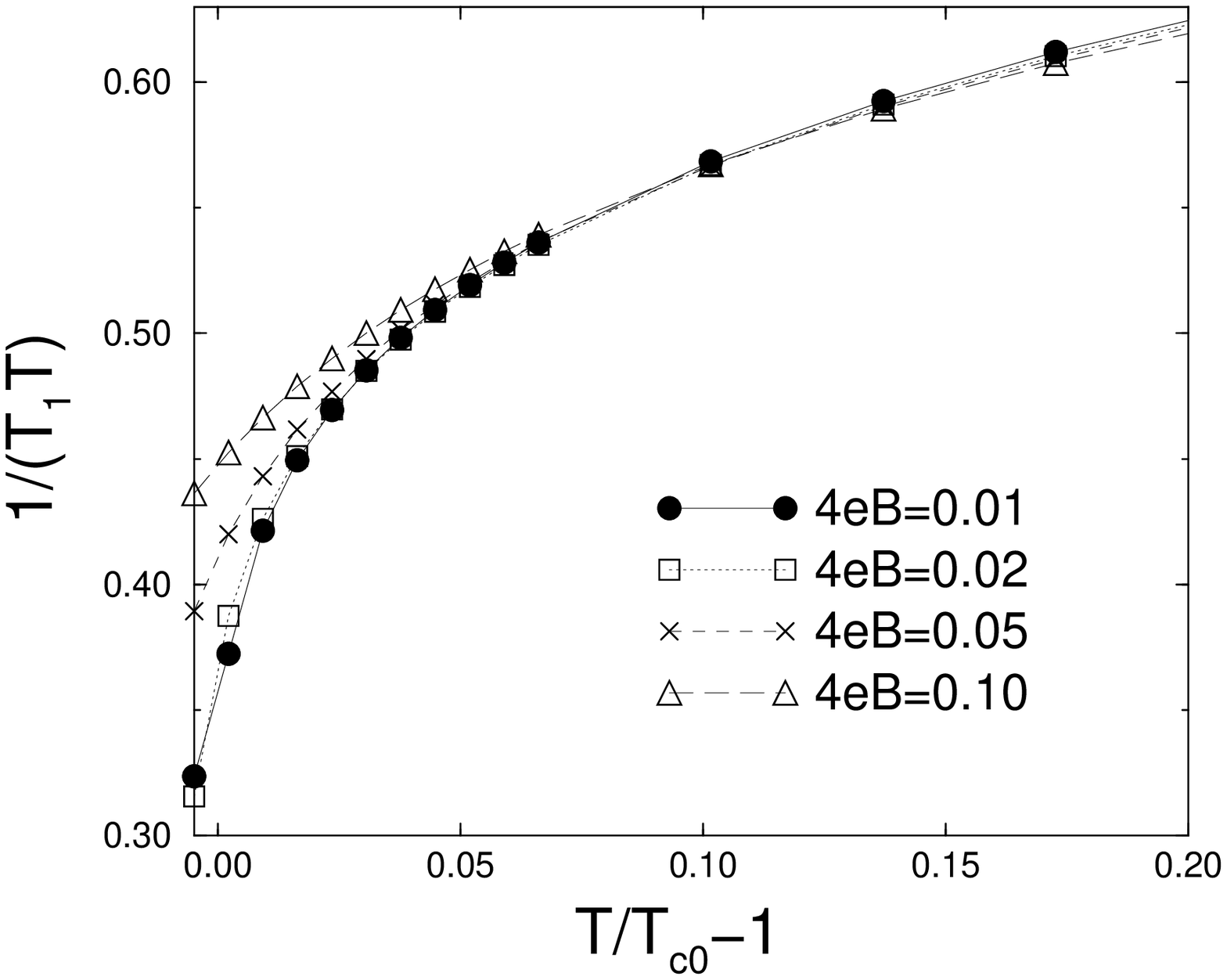}$$
     \caption{The calculated results for $1/T_{1}T$ under the magnetic field. 
             The magnetic field is the same as that in Fig.7. 
             The relatively strong pseudogap case $ g = -0.8 $.}
  \end{center}
\end{figure}

 We can see the different magnetic field dependences of the density of 
states according to the distance to the critical point (Fig.9). 
 The magnetic field effect is visible in the density of states just above 
$T_{{\rm c}}$ (Fig.9(a)). 
 The density of states at the low energy are recovered with increasing the 
magnetic field. 
On the other hand, the effect is almost invisible 
when the temperature is apart form $T_{{\rm c}}$ (Fig.9(b)).

\begin{figure}[ht]
  \begin{center}
   \epsfysize=6cm
$$\epsffile{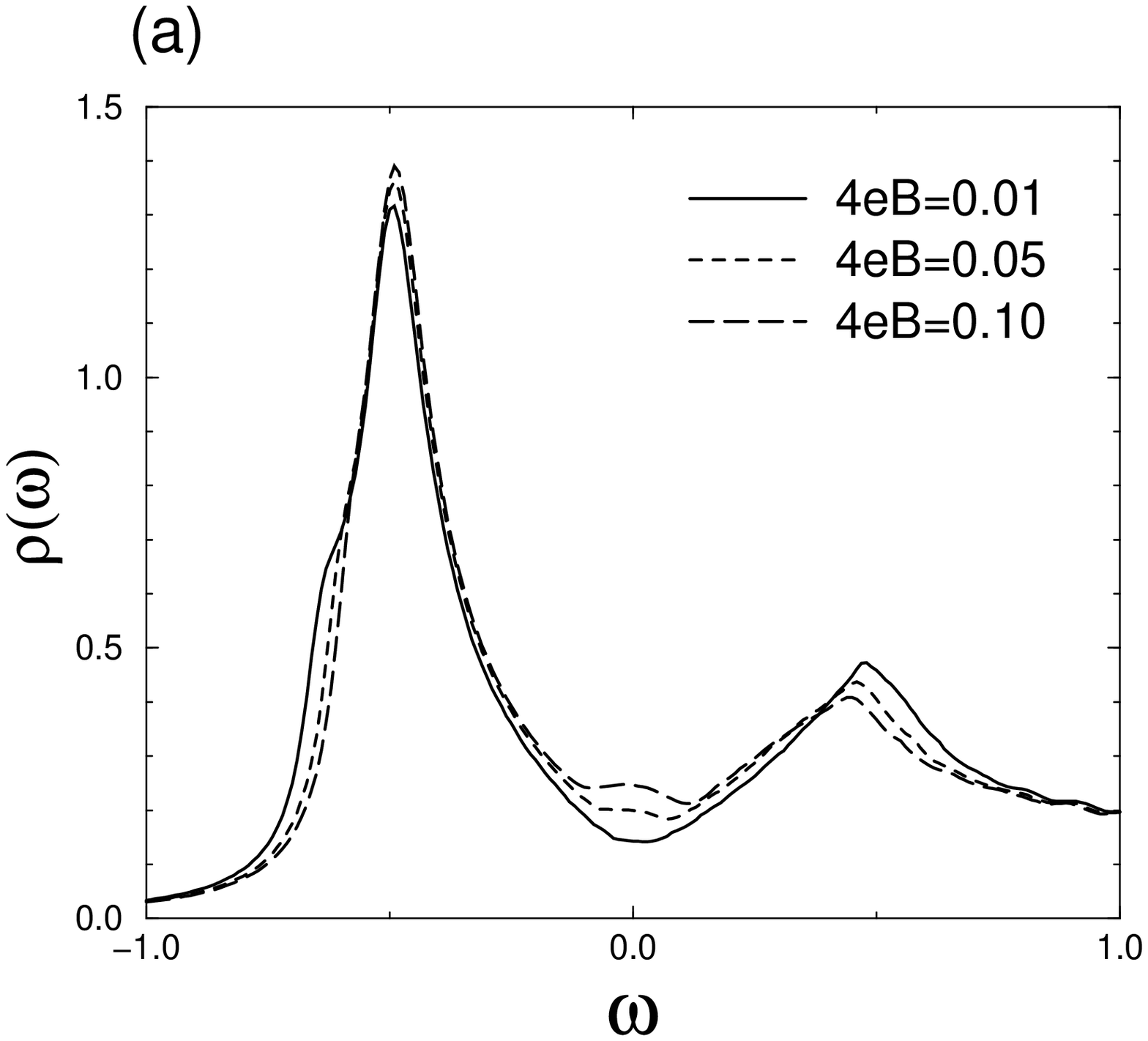}$$
   \epsfysize=6cm
$$\epsffile{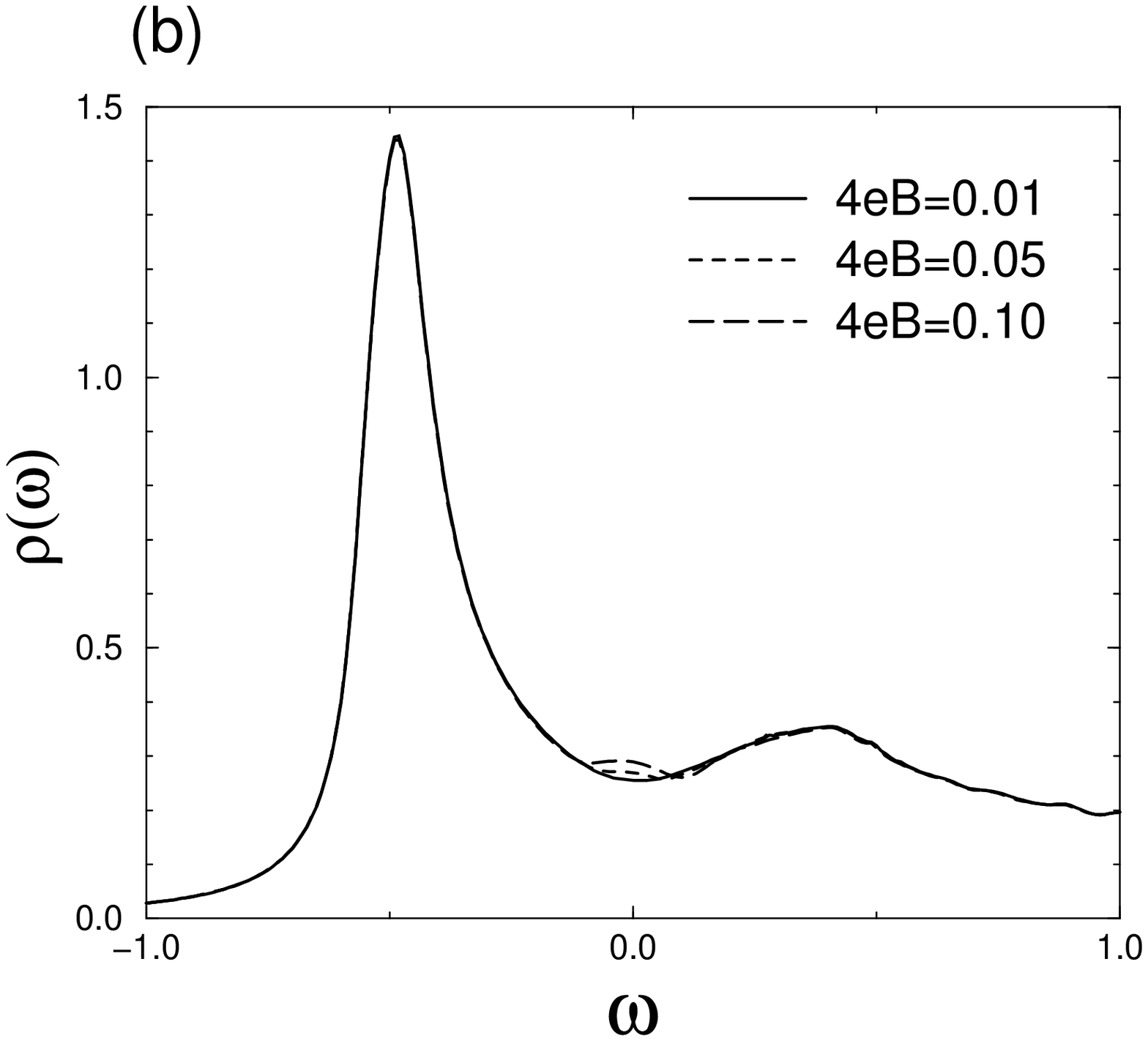}$$
     \caption{The magnetic field dependence of the density of states. 
              Here, $ g = -0.8 $ and $T_{{\rm c0}} = 0.1407$. 
              The magnetic field is varied as 
              $ 4 {\rm e} B = 0.01, 0.05, 0.1 $.
              (a) in the vicinity of $T_{{\rm c}}$, $T=0.141$ 
              (b) apart from $T_{{\rm c}}$, $T=0.15$ 
               }
  \end{center}
\end{figure}

 The results for the considerably strong coupling case $ g = -1.0 $ is shown 
in Fig.10. In this case, the strong pseudogap anomaly exists in the wide 
temperature region. The magnetic field effects become still smaller. 
 The magnetic field dependence is narrowly observed 
in the vicinity of $T_{{\rm c}}$.

\begin{figure}[ht]
  \begin{center}
   \epsfysize=6cm
$$\epsffile{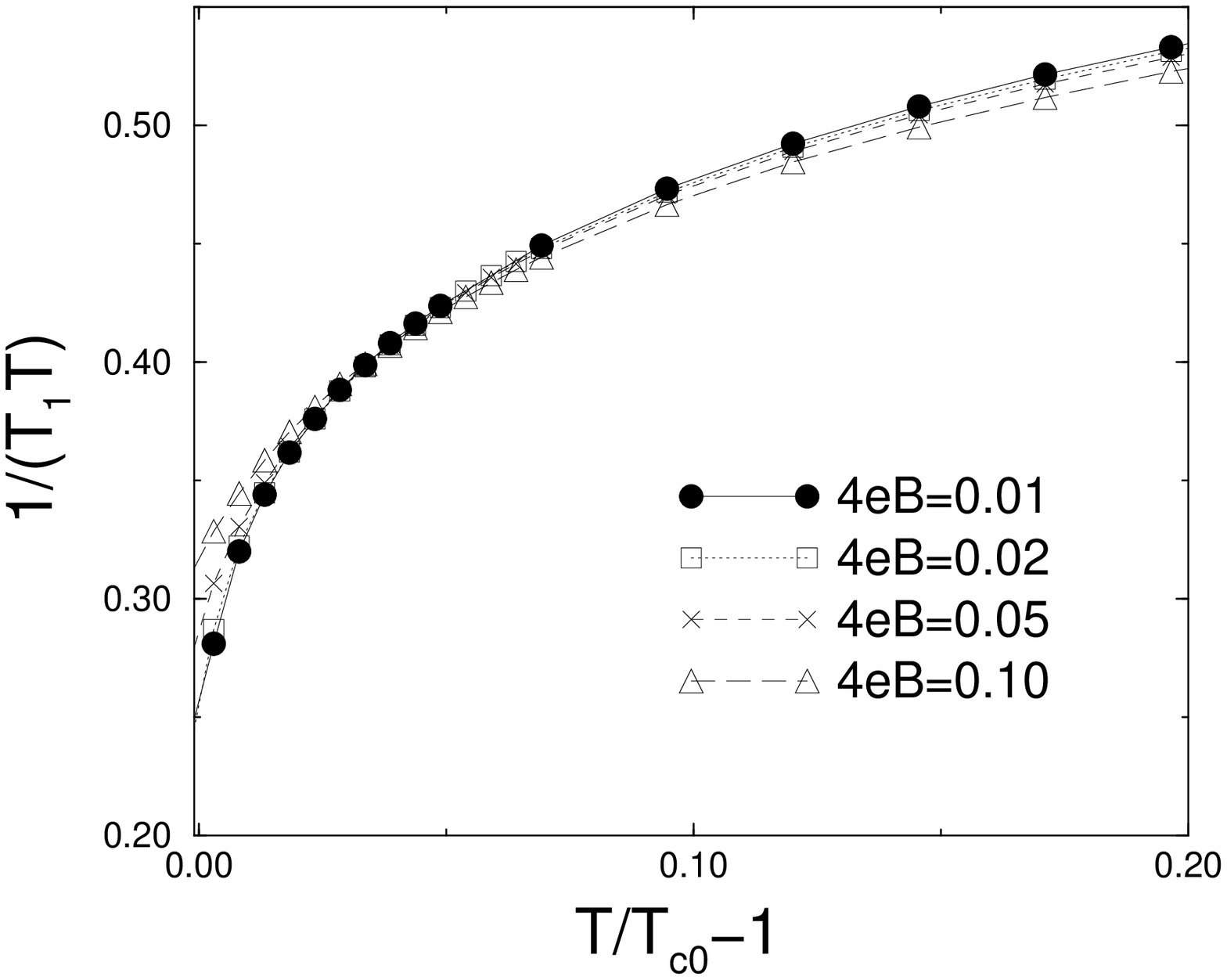}$$
     \caption{The calculated results for $1/T_{1}T$ under the magnetic field. 
             The magnetic field is the same as that in Fig.7. 
             The strong pseudogap case $ g = -1.0 $.}
  \end{center}
\end{figure}

 We consider that these strong coupling cases correspond to the under-doped 
cuprates. These behaviors well explain the experimental results in the 
under-doped cuprates.~\cite{rf:zheng} The weak effect in the vicinity of 
$T_{{\rm c}}$ is also observed in the experimental results. 
 Thus, the interpretation of the experimental results as a negative evidence 
for the pairing scenario is inappropriate. 

 The strength of the superconducting coupling is indicated by the ratio 
$T_{{\rm c}}/\varepsilon_{{\rm F}}$. The ratio increases due to the mass 
renormalization by the electron-electron correlation.  
 It should be considered that the mass-renormalization is enhanced 
with decreasing the doping quantity. 
 The attractive interaction becomes strong at the same time, 
since the anti-ferromagnetic spin fluctuations are enhanced. 
 Therefore, it is expected that the superconductivity  
becomes the strong coupling as the doping quantity decreases. 
 Thus, the strength of the superconducting coupling naturally changes with 
the doping in accordance with our expectation. 

 It should be noticed that the change of the magnetic field effects is 
continuous from weak to strong coupling. In other words, the calculated 
results explain the NMR measurements continuously and entirely from over-doped 
to under-doped cuprates. Therefore, the recent high field NMR measurements 
including their doping dependence are regarded as an affirmative evidence 
for the pairing scenario. 

 In order to confirm the effect of the Landau degeneracy to enhance the 
fluctuations, we show the Fig.11. 
 In Fig.11, the horizontal axis is scaled by the critical temperature under 
the magnetic field $ T_{{\rm cH}} $. 
 By keeping the distance to the critical point, we can remove the 
effect of the suppression of the superconductivity. Therefore, we can see the 
effect of the Landau degeneracy.

\begin{figure}[ht]
  \begin{center}
   \epsfysize=6cm
$$\epsffile{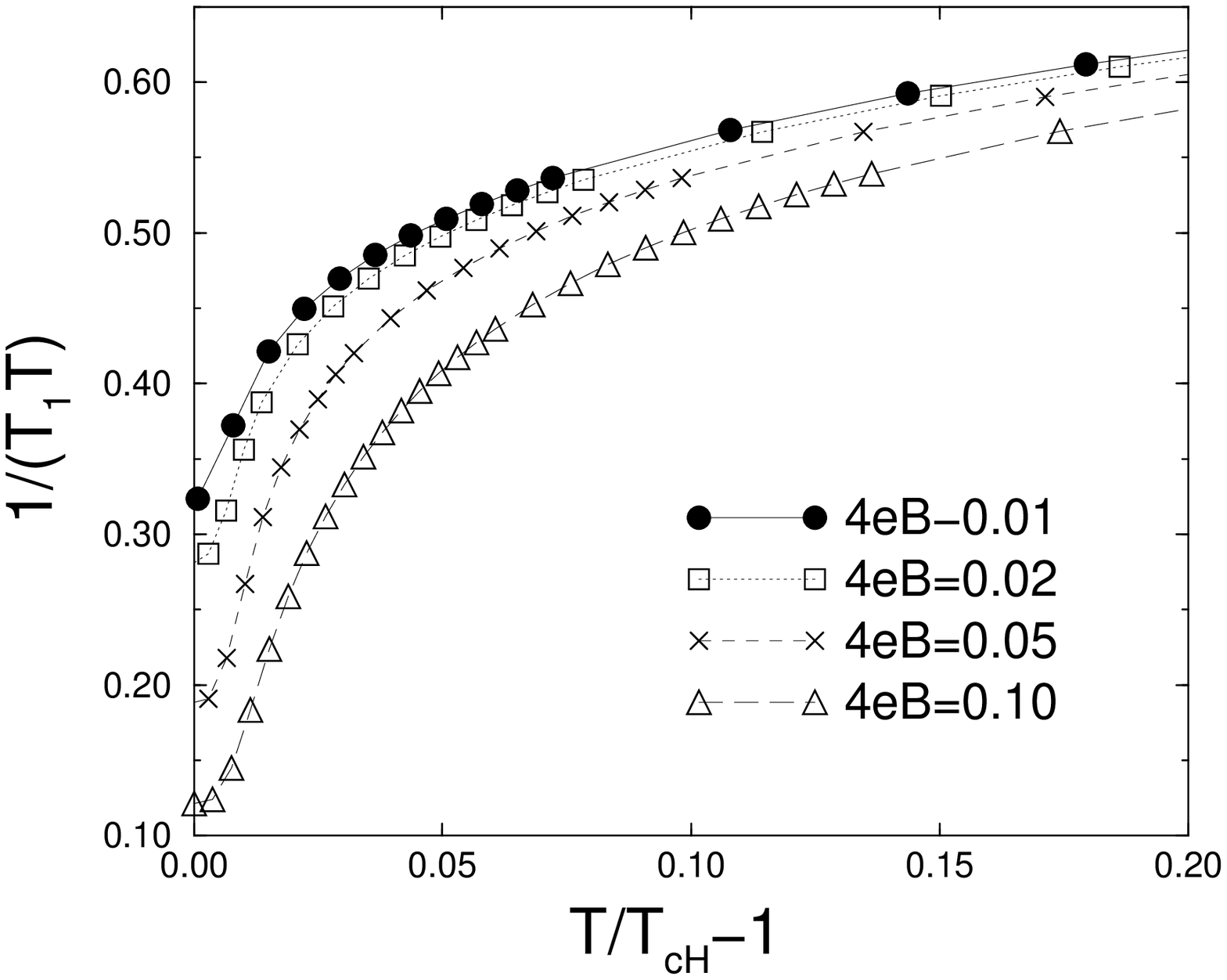}$$
     \caption{The results for $1/T_{1}T$ in which the temperature is scaled by 
              the critical temperature under the magnetic field $T_{{\rm cH}}$.
              The magnetic field is the same as that in Fig.7. 
              Here, $ g = -0.8 $.}
  \end{center}
\end{figure}

 The results show that $1/T_{1}T$ decreases with increasing the magnetic field.
It is because of the Landau degeneracy. The Landau degeneracy enhances the 
superconducting fluctuations and make the pseudogap stronger. Then, 
$1/T_{1}T$ is still more reduced. 
 Therefore, even in the rather weak pseudogap case, the pseudogap may be 
observed clearly under the high magnetic field. In other words, 
the magnetic field makes the pseudogap visible in more over-doped cuprates.

 At the last of this section, we consider the effect of the exchange 
enhancement. The exchange enhancement is taken into account within the 
random phase approximation (RPA). 
 The basic results about the magnetic field effects are not changed. 
However, it is definitely shown that the peak of $1/T_{1}T$ 
($T=T^{*}$) does not change in the strong coupling case, while the peak changes
in the weak coupling case. 
 The dynamical spin susceptibility 
$\chi_{{\rm RPA}}(\mbox{\boldmath$k$}, \omega)$ calculated by RPA is expressed 
as follows.

\begin{eqnarray}
  \chi_{{\rm RPA}}^{{\rm R}}(\mbox{\boldmath$q$}, \omega) & = & 
  \frac{\chi_{0}^{{\rm R}}(\mbox{\boldmath$q$}, \omega)}
       {1 - U \chi_{0}^{{\rm R}}(\mbox{\boldmath$q$}, \omega)},
\\
   \chi_{0}(\mbox{\boldmath$q$}, {\rm i} \omega_{n}) & = & 
   -T \sum_{\mbox{\boldmath$k$},\omega_{m}} 
     {\mit{\it G}} (\mbox{\boldmath$k$},{\rm i} \omega_{m}) 
      {\mit{\it G}} (\mbox{\boldmath$k$}+\mbox{\boldmath$q$},
                  {\rm i} \Omega_{m} + {\rm i} \omega_{n}).
\nonumber \\
\end{eqnarray}

 We fix $U=1.5$ afterward. 
 $1/T_{1}T$ is calculated by eq.(3.1). Here, we take into account the 
momentum dependence of the hyperfine coupling  
$ |A(\mbox{\boldmath$q$})|^{2} = \frac{1}{2} [\{A_{1} + 2 B (\cos(q_{x})+
\cos(q_{y}))\}^{2} + \{A_{2} + 2 B (\cos(q_{x})+\cos(q_{y}))\}^{2}] $. 
  The hyperfine coupling constants $A_{1}, A_{2}$ and $B$ is evaluated as 
$A_{1} = 0.84 B$ and $A_{2} = -4 B$.~\cite{rf:barzykin}
 The following results are not affected by the choice of the parameters, 
qualitatively. 

 The calculated results are shown in Figs.12 and 13. 
 In the high temperature region, $1/T_{1}T$ is enhanced owing to the exchange 
enhancement. 
 Near the critical temperature, $1/T_{1}T$ is reduced owing to the 
superconducting fluctuations. As a result, $1/T_{1}T$ shows its peak 
at $T=T^{*}$ above $T_{{\rm c}}$. It is a well-known pseudogap phenomenon  
in NMR measurements.

\begin{figure}[ht]
  \begin{center}
   \epsfysize=6cm
$$\epsffile{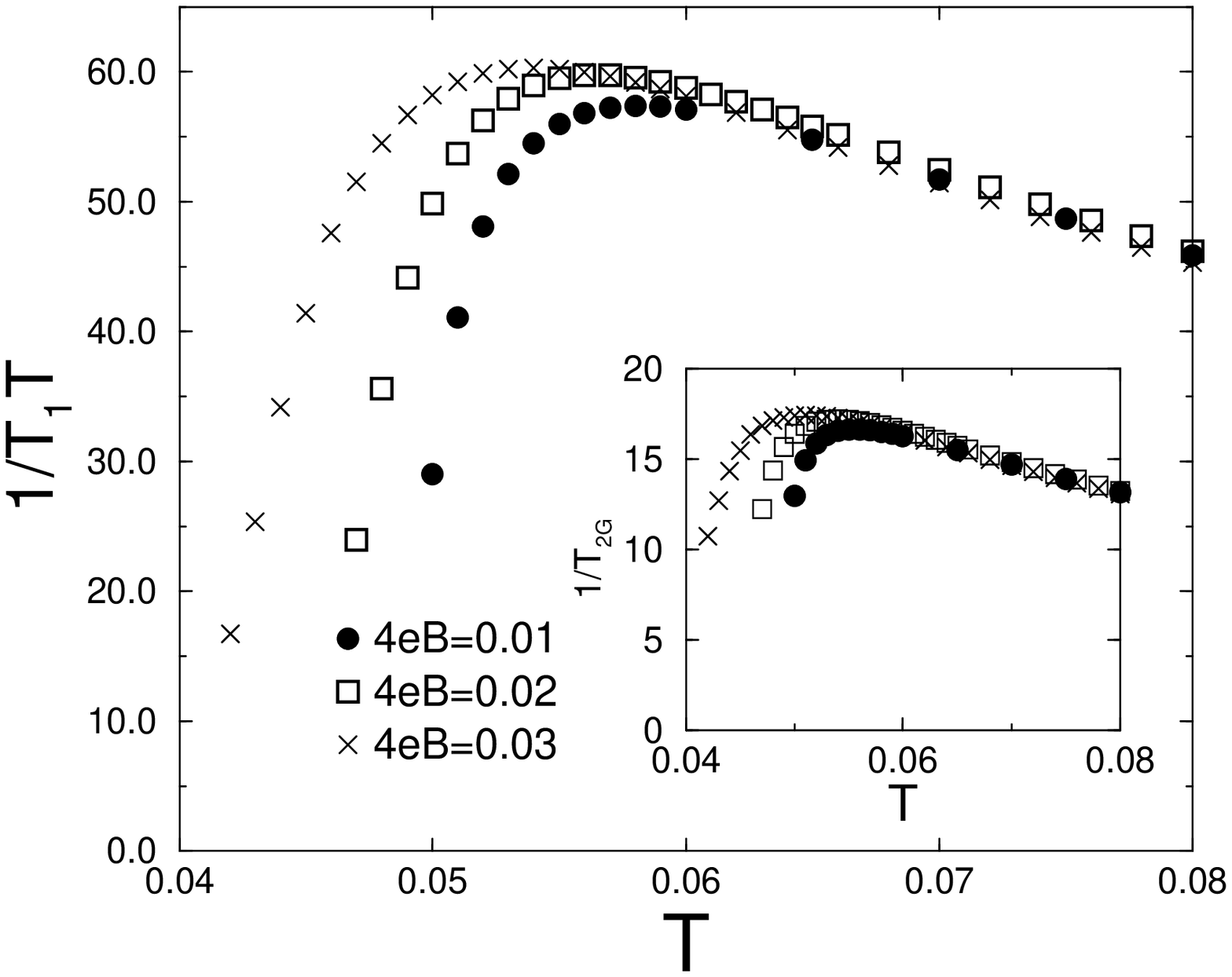}$$
     \caption{The results for $1/T_{1}T$ including the effects of the exchange 
              enhancement. The weak pseudogap case $g=-0.5$. 
              the magnetic field is varied as $ 4 {\rm e} B = 0.01, 0.02, $
              and $ 0.03 $ in our unit. The inset shows the results for
              $1/T_{2G}$}
  \end{center}
\end{figure}

\begin{figure}[ht]
  \begin{center}
   \epsfysize=6cm
$$\epsffile{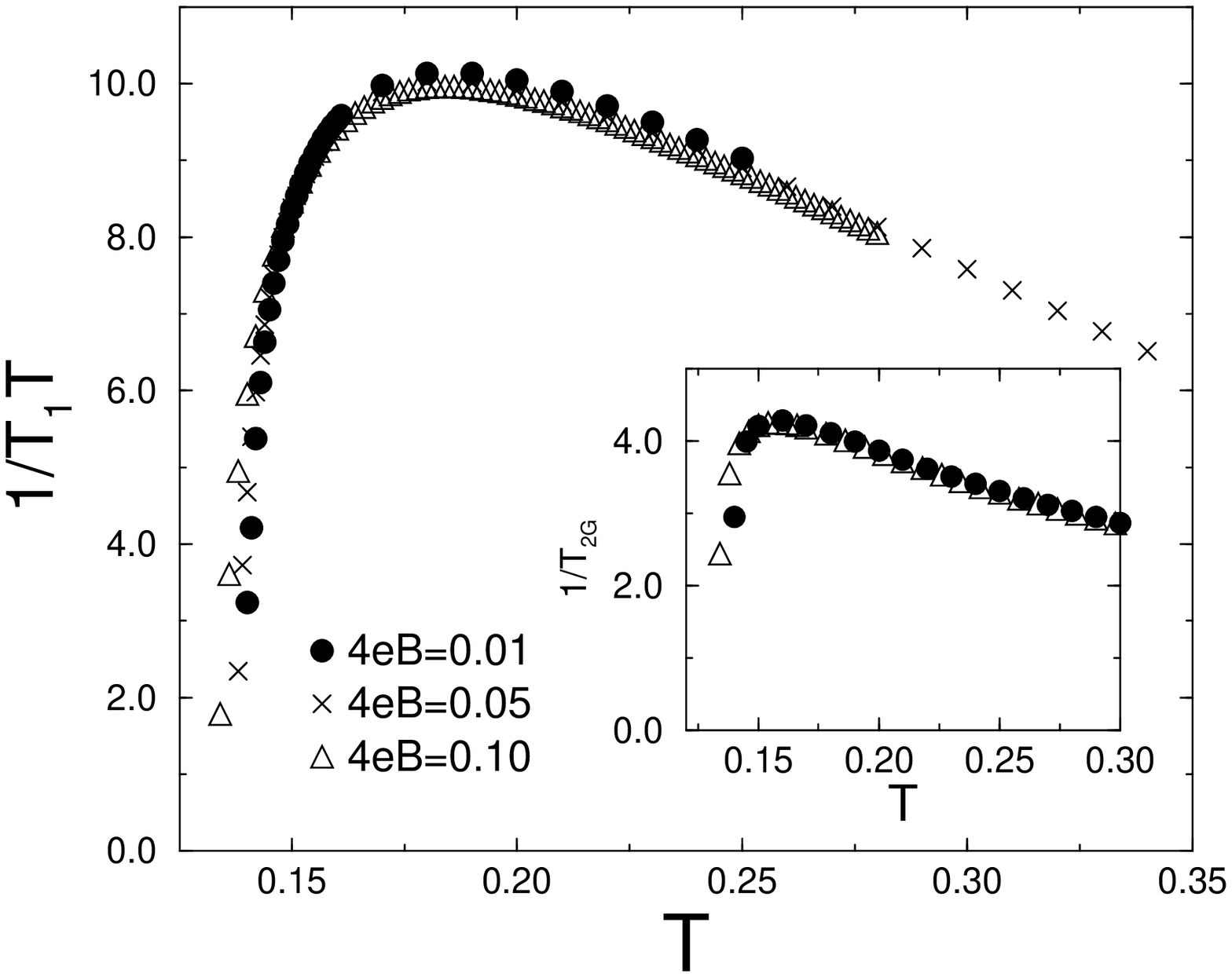}$$
     \caption{The results for $1/T_{1}T$ including the effects of the exchange 
              enhancement. The relatively strong pseudogap case $g=-0.8$. 
              the magnetic field is varied as $ 4 {\rm e} B = 0.01, 
              0.05,$ and $ 0.1 $ in our unit. The inset shows the results for
              $1/T_{2G}$} 
  \end{center}
\end{figure}

 In the weak coupling case $g=-0.5$ (Fig.12), the magnetic field effect is 
clearly observed. $T^{*}$ is lowered by the magnetic field. 
 On the other hand, in the relatively strong coupling case $g=-0.8$ (Fig.13), 
the magnetic filed effect is remarkably small. 
 $T^{*}$ is not changed by the magnetic field. 
$1/T_{1}T$ shows the magnetic filed dependence only in the vicinity of 
the critical temperature $T_{{\rm c}}$. 
 These features are the same as those derived by the calculation 
without the effect of the exchange enhancement. 

 The results for the spin-echo decay rate $1/T_{2G}$ are shown in the inset of 
Figs.12 and 13. $1/T_{2G}$ is calculated by the following expression. 

\begin{eqnarray}
  1/T_{2G}^{2}  =   \sum_{\mbox{\boldmath$q$}} 
                     [|A_{\parallel}(\mbox{\boldmath$q$})|^{2} 
        {\rm Re} \chi_{{\rm s}}^{{\rm R}} (\mbox{\boldmath$q$}, 0)]^{2}
                  -  [\sum_{\mbox{\boldmath$q$}} 
                      |A_{\parallel}(\mbox{\boldmath$q$})|^{2} 
        {\rm Re} \chi_{{\rm s}}^{{\rm R}} (\mbox{\boldmath$q$}, 0)]^{2}.
\end{eqnarray}

 Here, the dynamical spin susceptibility is calculated by RPA, and 
$|A_{\parallel}(\mbox{\boldmath$q$})|^{2} = 
\{A_{2} + 2 B (\cos(q_{x})+\cos(q_{y}))\}^{2}$.~\cite{rf:barzykin}
 $1/T_{2G}$ shows the pseudogap phenomena. However, the effect of the pseudogap
on $1/T_{2G}$ is weaker than that on $1/T_{1}T$. The pseudogap appears in the 
narrower temperature region. $1/T_{2G}$ shows its peak below the pseudogap 
onset temperature $T^{*}$ in $1/T_{1}T$. 
 These results are consistent with the experimental 
results.~\cite{rf:tokunaga} 

 These results indicate that the effects of the pseudogap are weak 
on the real part of the spin susceptibility rather than on the imaginary part
at the low frequency. 
 The dissipation (imaginary part) directly reflects the low energy density of 
state. However, the static property (real part) does not necessarily so.  
 In other wards, the pseudogap suppresses the weight of the spin 
susceptibility at low frequency. However, the effect on the total weight is 
rather small.  
 In particular, the $d$-wave pseudogap only weakly affects the real part 
near the anti-ferromagnetic wave vector $\mbox{\boldmath$q$} = (\pi,\pi)$. 
 The momentum dependence of the hyperfine coupling 
$A_{\parallel}(\mbox{\boldmath$q$})$ 
still more weaken the effect of the pseudogap on $1/T_{2G}$. 
 The above features are in common with the superconducting 
state.~\cite{rf:bulut}
 That is natural because the pseudogap and the superconducting gap have 
the same $d_{x^{2}-y^{2}}$-wave form. 
 The magnetic field dependence of $1/T_{2G}$ has the same features as those 
of $1/T_{1}T$.

\section{Transport in the Pseudogap Phase}

 In the previous sections, we have paid attention to the NMR spin-lattice 
relaxation rate $1/T_{1}T$ and its magnetic field dependence in the pseudogap 
phase. 
 Besides that, the pseudogap affects many other measurements. These effects 
may be understood by considering the suppression of the one-particle spectral 
weight and that of the low frequency anti-ferromagnetic spin 
fluctuations.~\cite{rf:yanasepg} 

 In particular, the transport phenomena are enough interesting to be discussed 
here, because they reflect the characteristic momentum dependences of 
High-$T_{{\rm c}}$ cuprates and the relationship between the spin fluctuations 
and the superconducting fluctuations. 
 The following qualitative discussion deserves to be described here, 
because there is no explicit calculation based on our understanding. 

 First, we describe how the transport phenomena in under-doped cuprates 
are understood in the normal phase ($T > T^{*}$). 
 They are anomalous at a glance. However, we can understand them by 
considering the magnetic interaction caused by the anti-ferromagnetic 
spin fluctuations.~\cite{rf:yanasetr,rf:kontani} 
 The momentum dependence of the lifetime of quasi-particles is important to 
understand the transport properties. 
 The momentum dependent lifetime is due to the scattering by the 
anti-ferromagnetic spin fluctuations.~\cite{rf:stojkovic,rf:yanasetr}

 'Hot spot' means the part of the Fermi surface in which 
$\varepsilon_{\mbox{\boldmath$k$}} = 
\varepsilon_{\mbox{\boldmath$k$}+\mbox{\boldmath$Q$}} $. 
Here, $\mbox{\boldmath$Q$}$ is a anti-ferromagnetic wave vector 
$\mbox{\boldmath$Q$} = (\pi,\pi)$.
 At 'hot spot', quasi-particles suffer an immediate scattering by the 
anti-ferromagnetic spin fluctuations at $\mbox{\boldmath$q$} = 
\mbox{\boldmath$Q$}$. 
 'Cold spot' is the area on the Fermi surface far from 'hot spot'. 
 There, quasi-particles do not suffer the immediate scattering. 
 Therefore, the lifetime is long at 'cold spot' and short at 'hot spot'. 

 This momentum dependent lifetime is a general property of the systems with 
anti-ferromagnetic spin fluctuations.~\cite{rf:rosch} 
 For High-$T_{{\rm c}}$ cuprates, 'hot spot' is located near $(\pi,0)$ and 
its symmetric points. 'Cold spot' is located near $(\pi/2,\pi/2)$. 
 At the same time, the pseudogap is large at 'hot spot' and small at 
'cold spot' because of its $d_{x^{2}-y^{2}}$-wave shape. 

 'Hot spot' does not contribute to the in-plane conductivity, because the 
conductivity is almost determined by the most easily flowing quasiparticles. 
 The in-plane conductivity is mainly determined by 'cold spot'. 
 The quasiparticles at 'cold spot' are sure to have the $T^{2}$-damping 
rate at the low temperature limit which is consistent with the conventional 
Fermi liquid theory. 
 However, they have the $T$-linear damping rate above the relatively low 
crossover temperature ($T > T_{{\rm cr}}$). 
 It is because of the low energy magnetic excitations. 
 The transformation of the Fermi surface which leads to a form more 
appropriate to the nesting reduces the crossover temperature $T_{{\rm cr}}$. 
 The transformation itself is due to the anti-ferromagnetic spin 
fluctuations.~\cite{rf:yanasetr} 
 As a result, the in-plane resistivity shows a $T$-linear law in the normal 
phase ($T > T^{*} >T_{{\rm cr}}$). 
 It should be noticed that $T$-linear resistivity is not due to the 
Curie-Weiss law $ \chi_{{\rm s}}(\mbox{\boldmath$Q$}) \propto 1/(T+\theta) $, 
or $ 1/T_{1}T \propto 1/(T+\theta) $. 
 The calculations inappropriately treating the momentum 
dependent lifetime attribute the $T$-linear resistivity to the Curie-Weiss 
law.~\cite{rf:moriya,rf:kohno2} For example, the approximate relation between 
the in-plane resistivity $\rho_{{\rm ab}}$ and $1/T_{1}T$, 
$\rho_{{\rm ab}} \propto T^{2}/(T_{1}T)$ is derived.~\cite{rf:kohno2} 
 If appropriately considering 'hot spot' and 'cold spot', the $T$-linear 
resistivity is realized more generally, but in more high temperature 
region.~\cite{rf:yanasetr} 
 This generality is important to understand the $T$-linear in-plane 
resistivity in the pseudogap phase. 

 The other important character of High-$T_{{\rm c}}$ cuprates is a 
momentum dependence of the interlayer hopping matrix element 
$t_{\perp}(\mbox{\boldmath$k$})$. 
 The band calculation has shown that the dispersion along the {\it c}-axis is 
large at 'hot spot' and is nearly $0$ at 'cold spot'.~\cite{rf:okanderson} 
 $t_{\perp}(\mbox{\boldmath$k$})$ is approximately expressed 
as~\cite{rf:okanderson,rf:ioffe} 

\begin{eqnarray}
  \label{eq:c-hopping}
  t_{\perp}(\mbox{\boldmath$k$}) \propto 
  (\frac{{\rm cos} k_{x} - {\rm cos} k_{y}}{2})^{2}. 
\end{eqnarray}

 Since the quasiparticle velocity along the {\it c}-axis is nearly $0$ 
at 'cold spot', 'cold spot' 
does not contribute to the {\it c}-axis conductivity. On the other hand, 
the contribution from 'hot spot' is reduced by the short lifetime, in spite of 
the large velocity along the {\it c}-axis. 
 As a result, the {\it c}-axis transport becomes incoherent. 
 Thus, we can understand the coherent in-plane conductance and the incoherent 
{\it c}-axis conductance at the same time.~\cite{rf:yanasetr}

 The momentum dependent lifetime enhances the Hall coefficient 
$ R = \sigma_{xy}/\sigma_{xx}^{2}H $.~\cite{rf:yanasetr} 
 However, the vertex correction plays a more important role for the Hall 
coefficient.~\cite{rf:kontani} 
 It is because of the momentum derivative of the total current $J_{\nu}$ in 
the general formula given by Kohno and Yamada.~\cite{rf:kohno} 
 The Hall coefficient is strongly enhanced by the vertex 
correction.
 In the conventional metals, the vertex correction gives only an constant 
factor arising from the Umklapp scattering and has no significant 
effect.~\cite{rf:yamada} The significance of the vertex correction is also due 
to the anti-ferromagnetic spin fluctuations. 
 The vertex correction is not so important for the longitudinal conductivity 
even in the systems with spin fluctuations.~\cite{rf:kontani} 

 Here, we consider the transport phenomena in the pseudogap phase. 
 The main effects of the pseudogap on the transport phenomena are the 
following two points. One is the pseudogap itself. The other is the 
suppression of the anti-ferromagnetic spin fluctuations. 

 Because of the singlet pairing correlation, the low frequency part of the 
anti-ferromagnetic spin fluctuations is expected to be suppressed. 
 Indeed, NMR measurements show the suppression of $ (T_{1}T)^{-1} $ and 
$ (T_{2G})^{-1} $.~\cite{rf:NMR,rf:tokunaga,rf:itoh} 
 The low frequency part of the anti-ferromagnetic spin fluctuations causes 
the quasiparticle damping. 
 Therefore, the quasiparticle damping due to the anti-ferromagnetic spin 
fluctuations is immediately affected by the pseudogap.

 As is shown in \S2, the imaginary part of the self-energy 
due to the superconducting fluctuations, 
${\rm Im} {\mit{\it \Sigma}}^{{\rm R}} (\mbox{\boldmath$k$}, \omega)$
is remarkably large in the pseudogap phase. 
 The large imaginary part leads to the pseudogap near $(\pi,0)$. 
 Therefore, the contribution to the conductivity from the quasiparticles near 
$(\pi,0)$ is remarkably suppressed by the pseudogap itself. 
However, quasiparticles at 'hot spot' 
do not contribute to the in-plane conductivity from the beginning. 
 The in-plane conductivity is determined by the contribution from 'cold spot'. 
 Therefore, the pseudogap itself is not important for the in-plane 
conductivity. 
 The suppression of the anti-ferromagnetic spin fluctuations slightly 
reduces the scattering at 'cold spot'. Quasiparticles at 'cold spot' are not 
affected by the strong scattering due to the spin fluctuations 
at $\mbox{\boldmath$q$} = \mbox{\boldmath$Q$}$. 
 Therefore, the effect of the suppression of the anti-ferromagnetic spin 
fluctuations is small at 'cold spot' rather than at 'hot spot'. 
 As a result, the in-plane resistivity slightly deviates downward and 
keep the $T$-linear law.~\cite{rf:yanasetr} 
This behavior is observed in many under-doped 
compounds and the downward deviation coincides with the 
pseudogap.~\cite{rf:transport} 
 Thus, the $T$-linear resistivity generally appears owing to the low frequency 
magnetic excitations. 
 The $T$-linear law in the pseudogap phase can not be understood by the 
phenomenological relation, $\rho_{{\rm ab}} \propto 
T^{2}\chi_{{\rm s}}(\mbox{\boldmath$Q$})$ or 
$\rho_{{\rm ab}} \propto T^{2}/(T_{1}T)$. 

 On the other hand, the pseudogap itself has more drastic effect on the 
{\it c}-axis conductivity. The {\it c}-axis conductance is determined by the 
contribution from the vicinity of $(\pi,0)$. 
 Quasiparticles near $(\pi,0)$ decrease the contribution to
the {\it c}-axis conductivity owing to the pseudogap. 
The pseudogap is large there. 
 Therefore, the {\it c}-axis conductivity is remarkably suppressed 
by the pseudogap. 
 This effect is confirmed within the formalism in this paper. 
 We calculate the {\it c}-axis resistivity by using the Kubo formula and 
neglecting the vertex correction. 
 The {\it c}-axis conductivity is expressed as 

\begin{eqnarray}
  \sigma_{{\rm c}}(T)  =  d  e^{2} 
     \sum_{\mbox{\boldmath$k$}} t_{\perp}^{2}(\mbox{\boldmath$k$})
     \int \frac{{\rm d} \omega}{\pi} 
     (-f'(\omega))
 {\rm Im} {\mit{\it G}}^{{\rm R}}(\mbox{\boldmath$k$},\omega)
      {\rm Im} {\mit{\it G}}^{{\rm R}}(\mbox{\boldmath$k$},\omega). 
\end{eqnarray}

 Here, $d$ is the interlayer distance.
 We normalize the conductivity by the constant factor $d e^{2}$. 
 The calculated result is shown in Fig.14.
 The {\it c}-axis resistivity shows the semi-conductive behavior near the 
critical point $T_{{\rm c}}$. 
 It is because the scattering due to the superconducting fluctuations 
becomes remarkable with approaching the critical temperature.

\begin{figure}[ht]
  \begin{center}
   \epsfysize=5cm
$$\epsffile{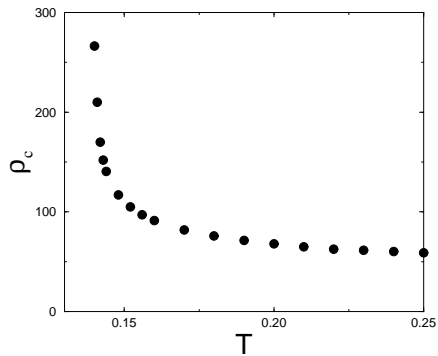}$$
     \caption{The normal state c-axis resistivity $\rho_{{\rm c}}$. 
              Here, $g=-0.8$. 
              }
  \end{center}
\end{figure}

 Thus, we can understand the drastic increase of the {\it c}-axis resistivity 
in the pseudogap phase, while the in-plane resistivity changes only a little. 

 Because of the momentum dependence of the hopping matrix element, 
the {\it c}-axis transport reflects the electronic state near $(\pi,0)$. 
 Therefore, we can see the pseudogap by observing the {\it c}-axis optical 
conductivity,~\cite{rf:homes} while the in-plane optical conductivity does not 
clearly indicate the pseudogap. 

 For the Hall conductivity, both two points play an important role. 
 Because of the suppression of the spin fluctuations, the enhancement of the 
Hall coefficient due to the spin fluctuations is reduced. 
 Moreover, since the vertex correction is large around 
'hot spot',~\cite{rf:kontani} the pseudogap itself 
affects the vertex correction. The pseudogap opens at 'hot spot' and reduces 
the effects of the vertex correction. 
 Due to the two effects, the Hall coefficient is reduced and shows its peak 
in the pseudogap phase. 

 These results well explain the observed transport phenomena in the pseudogap 
phase.~\cite{rf:transport}
 Thus, the transport phenomena in the pseudogap phase is naturally understood 
by considering the $d_{x^{2}-y^{2}}$-wave pseudogap.

\section{Summary and Discussion}

 In this paper, we have shown that the pairing scenario based on the strong 
coupling superconductivity well explains the effects of the magnetic field 
on the pseudogap phenomena in High-$T_{{\rm c}}$ cuprates. 
 
 We have shown in the previous paper~\cite{rf:yanasepg} that the pseudogap 
phenomena are properly described as a precursor of the superconductivity 
under the reasonable conditions. 
 In this paper, we have used the same formalism for calculating the 
single-particle self-energy and introduce the magnetic field effects 
thorough the Landau level quantization. We explicitly calculated the NMR 
spin-lattice relaxation rate $1/T_{1}T$ to compare the obtained results with 
the results of the recent high field NMR measurements. 

 The dominant effect of the Landau quantization is the suppression of the 
superconductivity, while the Landau degeneracy itself enhances the 
superconducting fluctuations.  
 From the simple discussion, we can see that the characteristic magnetic field 
is scaled as $ B_{{\rm ch}} \propto t_{0}/b = \xi_{{\rm GL}}^{-2}$. 
Actually, the calculated results support this behavior. 
 In the relatively weak coupling case, the weak pseudogap is observed in the 
narrow temperature region. 
 Then, the characteristic magnetic field is small and the magnetic field 
effects are visible. 
 On the other hand, in the strong coupling case where the pseudogap is 
observed in the wide temperature region, the characteristic magnetic field is 
large. In particular, it is remarkably large near the onset temperature 
$T^{*}$. Therefore $T^{*}$ is almost independent of the magnetic field. 
The magnetic field effects are visible only in the vicinity of the 
critical temperature 
$T_{{\rm c}}$. It should be noticed that the characteristic magnetic field 
$ B_{{\rm ch}} $ near $T^{*}$ is different from that near $T_{{\rm c}}$. 
 When the pseudogap phenomena take place at $T^{*}$, 
the superconducting correlation length $\xi_{{\rm GL}}$ is still short. 
Therefore, the pseudogap is not so affected by the magnetic field near 
$T^{*}$. 

 By considering that the effective Fermi energy $\varepsilon_{{\rm F}}$ 
decreases and the attractive interaction increases with decreasing the doping 
quantity, the calculated results well explain the high field NMR measurements 
including their doping dependence. The explanation is continuous 
from over-doped to under-doped cuprates. 

 There is an interpretation that the magnetic field independence of the 
pseudogap phenomena in under-doped cuprates is an evidence denying the 
pairing scenarios for the pseudogap.~\cite{rf:zheng} 
 However, the pairing scenario based on the strong coupling superconductivity 
naturally explains the experiments including their doping dependence. 

 Moreover, the continuous understanding in the phase diagram 
rather support the pairing scenario. 
 In the pseudogap phase, the self-energy correction due to the superconducting 
fluctuations is a common mechanism in reducing the density of states and 
$1/T_{1}T$. 
 Because the pseudogap phenomena continuously take place from slightly 
over-doped to under-doped cuprates, their magnetic field dependences 
should be continuously understood. The pseudogap becomes strong as the doping 
quantity decreases. 
The magnetic field dependence of the 
weak pseudogap case can be understood within the conventional weak coupling 
theory for the superconducting fluctuations.~\cite{rf:eschrig,rf:zhengprivate} 
Our theory is an extension of the theory. 
 This fact indicates the correctness of our description for the pseudogap 
phenomena in under-doped cuprates based on the strong coupling 
superconductivity. 

 On the other hand, it is not clear whether the magnetic origin 
may be consistent with 
the magnetic field dependence, especially in the weak pseudogap case. 
 It is because the magnetic exchange coupling $J$ is the order of 
$J \sim 1000 {\rm K}$ and the applied magnetic field is the order of 
$\mu_{{\rm B}} B \sim 10 {\rm K}$. 

 Moreover, we have discussed the transport phenomena in the pseudogap phase. 
Generally speaking, the transport phenomena in the normal phase are explained 
by the effects of the anti-ferromagnetic spin fluctuations. 
 We have shown that the transport coefficients in the pseudogap phase are 
naturally understood by considering the characteristic momentum dependences of 
both the spin- and the superconducting fluctuations 
in addition to the momentum dependence of the {\it c}-axis transfer matrix. 
The {\it c}-axis conductivity is mainly determined by the region near 
$(\pi,0)$. 
Therefore, the {\it c}-axis transport directly reflects the pseudogap. 
We have definitely shown the remarkable increase of the {\it c}-axis 
resistivity in the pseudogap phase. 

 Here, we give a brief discussion on the self-consistent calculation. 
In the self-consistent calculation the pseudogap is described in a similar 
way. The fundamental picture does not change also in the self-consistent 
calculation, although the renormalization effects on the TDGL parameters exist.
 However, the self-consistent T-matrix calculation is one of the methods 
introducing the criticality of the superconducting fluctuations. 
This effect corresponds to the forth order term in the Ginzburg-Landau 
description. 
 The forth order term in the Ginzburg-Landau action is expressed by the 
diagram shown in Fig.15. This term indicates the repulsive interaction 
between the fluctuating Cooper pairs (that is the mode coupling term). 

\begin{figure}[ht]
  \begin{center}
   \epsfysize=2.5cm
$$\epsffile{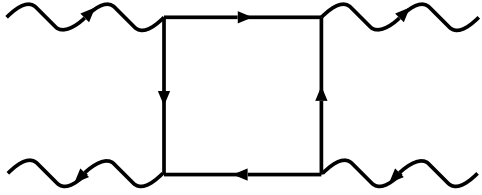}$$
     \caption{The diagram representing the repulsive interaction between the 
              fluctuating Cooper pairs. The wavy and solid lines represent 
              the propagator of the fluctuating Cooper pairs and that of the 
              fermions, respectively.} 
  \end{center}
\end{figure}

 The effect of this term is included in the self-consistent calculation 
at least in the level of the Hartree-Fock approximation. 
Thus, the criticality of the superconducting fluctuations is introduced.
 The criticality makes the magnetic field dependence still smaller. 
 To put it in detail, in the self-consistent calculation, $t_{0}$ depends on 
the magnetic field. As the magnetic field suppresses the pseudogap, 
$t_{0}(B)$ is reduced. Therefore, the distance to the superconductivity 
$t_{0}(B) + 2 b {\rm e} B$ varies more slowly than $t_{0}(0) + 2 b {\rm e} B$. 
 Thus, the magnetic field dependence is reduced by the criticality. 
 Anyway, the strong coupling superconductivity is the essential factor for the 
magnetic field independence, as we described in this paper. 
 The existence of the wide critical region is a result of the strong coupling 
superconductivity. 

 The more systematic measurements of the magnetic field dependences in the 
various doping rate will be an important verification to determine the origin 
of the pseudogap in High-$T_{{\rm c}}$ cuprates.

\section*{Acknowledgements}

 The authors are grateful to Mr. Koikegami for fruitful discussions. 
 The authors are grateful to Dr. G-q. Zheng for teaching us the experimental 
results. 
 Numerical computation in this work was partly carried out 
at the Yukawa Institute Computer Facility. 
 The present work was partly supported by a Grant-In-Aid for Scientific 
Research from the Ministry of Education, Science, Sports and Culture, Japan. 
 One of the authors (Y.Y) has been supported by a Research Fellowships of the 
Japan Society for the Promotion of Science for Young Scientists.


\begin{thebibliography}{99}
%
\bibitem{rf:bednortz} J. G. Bednortz and K. A. M$\ddot{{\rm u}}$ller: Z. Phys. 
B {\bf64} (1986) 189. 
\bibitem{rf:NMR} W. W. Warren, R. E. Walstedt, G. F. Brennert, R. J. Cava, 
R. Tycko, R. F. Bell and G. Dabbagh: Phys. Rev. Lett. {\bf62} (1989) 1193; 
 M. Takigawa, A. P. Reyes, P. C. Hammel, J. D. Thompson, R. H. Heffner, 
Z. Fisk and K. C. Ott: Phys. Rev. B {\bf43} (1991) 247; 
 M. H. Julien, P. Carretta, M. Horvati$\acute{{\rm c}}$: Phys. Rev. Lett. 
{\bf76} (1996) 4238; 
H. Yasuoka, S. Kambe, Y. Itoh and T. Machi: Physica. B {\bf199\&200} (1994) 
278; 
 K. Ishida, K. Yoshida, T. Mito, Y. Tokunaga, Y. Kitaoka, K. Asayama, 
Y. Nakayama, J. Shimoyama and K. Kishio: Phys. Rev. B {\bf 58} (1998) R5960. 
\bibitem{rf:homes} C. C. Homes, T. Timusk, R. Liang, D. A. Bonn and W. H. Hardy
: Phys. Rev. Lett. {\bf71} (1993) 1645;
D. N. Basov, R. Liang, B. Dabrowski, D. A. Bonn, W. N. Hardy and T. Timusk: 
Phys. Rev. Lett. {\bf77} (1996) 4090.
\bibitem{rf:transport} For example, T. Ito, K. Takenaka and S. Uchida: 
Phys. Rev. Lett. {\bf70} (1993) 3995; 
 K. Mizuhashi, K. Takenaka, Y. Fukuzumi and S. Uchida: Phys. Rev. B {\bf52} 
(1995) R3884;
 M. Oda, K. Hoya, R. Kubota, C. Manabe, N. Momono, T. Nakano and M. Ido: 
Physica. C {\bf281} (1997) 135. 
\bibitem{rf:ARPES} H. Ding, T. Yokoya, J. C. Campuzano, T. Takahashi, 
M. Randeria, M. R. Norman,  T. Mochiku, K. Kadowaki and J. Giapintzakis: 
Nature. {\bf382} (1996) 51; A. G. Loeser, Z. X. Shen, D. S. Dessau, 
D. S. Marshall, C. H. Park, P. Fournier and A. Kapitulnik: 
Science. {\bf273} 325. 
\bibitem{rf:renner} Ch. Renner, B. Revaz, J.-Y. Genoud, K. Kadowaki and 
\O. Fischer: Phys. Rev. Lett. {\bf80} (1998) 149. 
\bibitem{rf:momono} N. Momono, T. Matsuzaki, T. Nagata, M. Oda and M. Ido: 
to be published in J. Low. Temp. Phys. 
\bibitem{rf:yanasepg} Y. Yanase and K. Yamada: J. Phys. Soc. Jpn {\bf 68} 
(1999) 2999. 
\bibitem{rf:norman} M. R. Norman, H. Ding, M. Randeria, J. C. Campuzano, 
T. Yokoya, T. Takeuchi, T. Takahashi, T. Mochiku, K. Kadowaki, P. Guptasarma 
and D. G. Hinks: Nature. {\bf392} (1998) 157.
\bibitem{rf:tanamoto} T. Tanamoto, H. Kohno and H. Fukuyama: J. Phys. Soc. Jpn 
{\bf 63} (1994) 2739. 
\bibitem{rf:SDW} A. Kampf and J. R. Schrieffer: Phys. Rev. B {\bf41} (1990) 
6399; 
 A. V. Chubukov, D. K. Morr and K. A. Shakhnovich: Philos. Mag. B {\bf74} 
(1996) 563; 
 D. Pines: Z. Phys. B {\bf103} (1997) 129;
 A. V. Chubukov and J. Schmalian: Phys. Rev. B {\bf57} (1998) 11085. 
\bibitem{rf:emery} V. J. Emery and S. A. Kivelson: Phys. Rev. Lett. {\bf74} 
(1995) 3253. 
\bibitem{rf:phase} M. Franz and A. J. Millis: Phys. Rev. B {\bf58} (1998) 
14572; H. J. Kwon and A. T. Dorsey: Phys. Rev. B {\bf59} (1999) 6438.
\bibitem{rf:randeriareview} M. Randeria: preprint. (cond-mat/9710223); 
C. A. R. S$\acute{{\rm a}}$ de Melo, M. Randeria and J. R. Engelbrecht:
Phys. Rev. Lett. {\bf71} (1993) 3202. 
\bibitem{rf:Nozieres} P. Nozi$\grave{{\rm e}}$res and S. Schmitt-Rink: 
J. Low Temp. Phys. {\bf 59} (1985) 195; 
 A. J. Leggett: {\it Modern Trends in the Theory of Condensed Matter} 
ed. A. Pekalski and R. Przystawa (Springer-Verlag, Berlin, 1980). 
\bibitem{rf:resonance} O. Tchernyshyov: Phys. Rev. B {\bf56} (1997) 3372. 
\bibitem{rf:janko} B. Jank$\acute{{\rm o}}$, J. Maly and K. Levin: 
Phys. Rev. B {\bf56} (1997) 11407; 
 J. Maly, B. Jank$\acute{{\rm o}}$ and K. Levin: preprint. (cond-mat/9805018)
\bibitem{rf:zheng} G-q. Zheng, W. G. Clark, Y. Kitaoka, K. Asayama, K. Kodama, 
P. Kuhns and W. G. Moulton: Phys. Rev. B {\bf 60} (1999) R9947. 
\bibitem{rf:gorny} K. Gorny, O. M. Vyaselev, J. A. Martindale, V. A. Nandor, 
C. H. Pennington, P. C. Hammel, W. L. Hults, J. L. Smith, P. L. Kuhns, 
A. P. Reyes and W. G. Moulton:  Phys. Rev. Lett. {\bf82} (1999) 177. 
\bibitem{rf:mitrovic} V. F. Mitrovi$\acute{{\rm c}}$, H. N. Bachman, 
W. P. Halperin, M. Eschrig, J. A. Sauls, A. P. Reyes, P. Kuhns and  
W. G. Moulton: Phys. Rev. Lett. {\bf82} (1999) 2784; 
 V. F. Mitrovi$\acute{{\rm c}}$, H. N. Bachman, W. P. Halperin, M. Eschrig 
and J. A. Sauls: preprint. (cond-mat/9901232) 
\bibitem{rf:zhengprivate} G-q. Zheng: private communication. 
\bibitem{rf:eschrig} M. Eschrig, D. Rainer and J. A. Sauls: Phys. Rev. B 
{\bf 59} (1999) 12095. 
\bibitem{rf:monthoux} P. Monthoux, A. V. Balatsky and D. Pines: 
Phys. Rev. B {\bf 46} (1992) 14803. 
\bibitem{rf:moriya} T. Moriya, Y. Takahashi and K. Ueda: 
J. Phys. Soc. Jpn {\bf 59} (1990) 2905. 
\bibitem{rf:stojkovic} B. P. Stojkovi$\acute{{\rm c}}$ and D. Pines: Phys. 
Rev. Lett. {\bf76} (1996) 811; 
 B. P. Stojkovi$\acute{{\rm c}}$ and D. Pines: Phys. Rev. B {\bf55} (1997) 
8576. 
\bibitem{rf:yanasetr} Y. Yanase and K. Yamada: J. Phys. Soc. Jpn {\bf 68} 
(1999) 548. 
\bibitem{rf:dahm} T. Dahm, D. Manske and L. Tewordt: Phys. Rev. B {\bf55} 
(1997) 15274.
\bibitem{rf:koikegamisuper} S. Koikegami and K. Yamada: to be published in 
J. Phys. Soc. Jpn.  
\bibitem{rf:AL} L. G. Aslamazov and A. I. Larkin: Fiz. Tverd. Tela. {\bf 10} 
(1968) 1104. [Sov. Phys. Solid State {\bf 10} (1968) 875.]
\bibitem{rf:MT} K. Maki: Prog. Theor. Phys {\bf 40} (1968) 193.;
R. S. Thompson:  Phys. Rev. B {\bf 1} (1970) 327. 
\bibitem{rf:DOS} K. Kuboki and H. Fukuyama: J. Phys. Soc. Jpn {\bf 58} 
(1989) 376; 
 J. Heym: J. Low Temp. Phys. {\bf 89} (1992) 869; 
 M. Randeria and A. A. Varlamov: Phys. Rev. B {\bf 50} (1994) 10401; 
 C. Di Castro, R. Raimondi, C. Castellani and A. A. Varlamov: 
Phys. Rev. B {\bf 42} (1990) 10221; 
 A. A. Varlamov, G. Balestrino, E. Milani and D. V. Livanov: Adv. Phys. 
{\bf 48} (1999) 655. 
\bibitem{rf:ebisawa} H. Ebisawa and H. Fukuyama: Prog. Theor. Phys {\bf 46} 
(1971) 1042.
\bibitem{rf:yanasepro} Y. Yanase and K. Yamada: to be published in Physica. B 
(2000). 
\bibitem{rf:FFLO} P. Fulde and R. A. Ferrell: Phys. Rev. {\bf 135} (1964) A550;
 A. I. Larkin and Y. N. Ovchinnikov: Sov. Phys. -JETP {\bf 20} (1965) 762. 
\bibitem{rf:barzykin} V. Barzykin and D. Pines: Phys. Rev. B {\bf52} (1995) 
13585.
\bibitem{rf:tokunaga} Y. Tokunaga {\it et al.}: to be published in Physica. B 
(2000). 
\bibitem{rf:bulut} N. Bulut and D. J. Scalapino: 
Phys. Rev. Lett. {\bf67} (1991) 2898. 
\bibitem{rf:kontani} H. Kontani, K. Kanki and K. Ueda: Phys. Rev. B {\bf59} 
(1999) 14723; K. Kanki and H. Kontani: J. Phys. Soc. Jpn {\bf 58} (1999) 1614.
\bibitem{rf:rosch} A. Rosch: Phys. Rev. Lett. {\bf82} (1999) 4280; 
 A. Rosch: preprint. (cond-mat/9908245) 
\bibitem{rf:kohno2} H. Kohno and K. Yamada: Prog. Theor. Phys. {\bf85} (1991) 
13. 
\bibitem{rf:okanderson} O. K. Anderson, A. I. Liechtenstein, O. Jepsen and 
F. Paulsen: J. Phys. Chem. Solids. {\bf56} (1995) 1573. 
\bibitem{rf:ioffe} L. B. Ioffe and A. J. Millis: Phys. Rev. B {\bf58} (1998) 
11631. 
\bibitem{rf:kohno} H. Kohno and K. Yamada: Prog. Theor. Phys. {\bf80} (1988) 
623. 
\bibitem{rf:yamada} K. Yamada and K. Yosida: Prog. Theor. Phys. {\bf76} (1986) 
621. 
\bibitem{rf:itoh} Y. Itoh, T. Machi, S. Adachi, A. Fukuoka, K. Tanabe and 
H. Yasuoka: J. Phys. Soc. Jpn. {\bf67} (1998) 312. 
%
\end{thebibliography}
\end{document}